\documentclass[superscriptaddress,prc,amsfonts,aps,nofootinbib,notitlepage,10pt]{revtex4-1}
\usepackage{graphicx}
\usepackage{color}
\usepackage{amsmath}
\begin{document}

\title{\textit{Ab-initio} no-core shell model study of $^{18-24}$Ne isotopes}

\author{Chandan Sarma and Praveen C. Srivastava\footnote{Corresponding author: praveen.srivastava@ph.iitr.ac.in}}
\address{Department of Physics, Indian Institute of Technology Roorkee, Roorkee 247667, India}

\date{\hfill \today}


\begin{abstract}
	
We report \textit{ab initio} no-core shell model (NCSM) study of $^{18-24}$Ne isotopes for energy spectra, electromagnetic properties, and point-proton radii using three realistic $NN$ interactions. We have used inside nonlocal outside Yukawa (INOY), charge-dependent Bonn 2000 (CDB2K) and the chiral next-to-next-to-next-to-leading order (N$^3$LO) interactions. We are able to reach basis size up to $N_{max}$ = 6 for $^{18}$Ne and $N_{max}$ = 4 for the  $^{19-24}$Ne isotopes with m-scheme dimensions up to 1.0 $\times$  $10^9$ in case of $^{24}$Ne. We observed better results for INOY interaction in terms of the binding energies of ground state (g.s.), and overall all the three interactions provide good agreement with the experimental low-energy spectra. Our results for reduced $M1$ transition strengths and magnetic moments are close to the experimental values. We found that for long-range observables such as the $E2$ transition strengths, the electric quadrupole moments, and the point-proton radii ($r_p$), we need higher $N_{max}$ calculations to obtain results comparable to the experimental data. We have observed almost 6 \% increment in the converged $r_p$ as we increase the model space from $N_{max}$ = 4 to $N_{max}$ = 6.

\end{abstract}


\maketitle
\section{Introduction}
\label{sect1}
Atomic nuclei are quantum many-body systems, and understanding nuclear properties from first principles is one of the main goal of low-energy nuclear physics. This motivation is fulfilled by the $ab$ $initio$ approaches \cite{NCSM_r2,Maris,Ragnar,CCEI}, which aim to describe different nuclear phenomena based on the fundamental interactions among the nucleons \cite{ab-initio2, ab-initio3}. The growth in computing power in recent times has opened the path to apply different \textit{ab inito} approaches to investigate nuclear structure properties \cite{ab-initio1}.
These approaches involve sophisticated methods to solve the non-relativistic many-particle Schr\"{o}dinger equation by taking nucleon-nucleon interaction as the input. The modern nuclear interactions are either derived from the meson exchange theory \cite{CDB2K3} or Quantum Chromodynamics (QCD) \cite{QCD} based on chiral effective field theory ($\chi$EFT) \cite{EFT1, EFT2, EFT3, EFT4}. Due to the complex nature of nuclear interaction, the application of \textit{ab initio} approaches to nuclear physics is a challenging task.

Being one of the \textit{ab initio} approaches, the no-core shell model (NCSM) \cite{NCSM_r2,NCSM_r1} is a fundamental approach in comparison to  the conventional nuclear shell model \cite{SM1, SM2, SM3, SM4, SM5, SM6}. The NCSM is able to explain different properties of nuclei in the lighter-mass region \cite{NCSM_p1, NCSM_p2, NCSM_p3, NCSM_p4, NCSM_p5, NCSM_p6, NCSM_p7, NCSM_p8}. In this paper our aim is to study $^{18-24}$Ne isotopes from the neutron shell closure at $N$ = 8 to $N$ = 14 for different nuclear properties. The neon isotopic chain exhibit novel structures, starting from the Borromean structure of $^{17}$Ne \cite{Ne0} which is considered to be a candidate for two-proton halo structure, clustering in the ground-state of $^{20}$Ne doubly open-shell isotope to the neutron-rich $^{30}$Ne around which the island of inversion occurs.
In the past, the study on $^{20-22}$Ne isotopes was carried out for level structures, and electromagnetic transitions within the shell-model approach \cite{Ne1, Ne2}. In Ref. \cite{Ne2*}, the isospin dependence of deformation is studied for carbon and neon isotopes. In this paper, the electromagnetic moments and transition strengths are calculated using the deformed Hartree-Fock (HF) model and compared with the standard shell model results.  
The significant contribution of the $^{16}$O + $\alpha$ cluster structure to the g.s. (0$^+$) of  $^{20}$Ne
is discussed extensively in Refs. \cite{Ne3, Ne4}. The antisymmetrized {\color{black} molecular} dynamics (AMD) method is used in Ref. \cite{Ne4} to study cluster structure of $^{20}$Ne along with $^{12}$C and $^{16}$O nuclei.
In Ref. \cite{Ne5}, two $ab$ $initio$ methods, namely the quasi-exact in-medium no-core shell model (IM-NCSM) and the projected generator coordinate method (PGCM), are used to study even neon isotopes from $^{18}$Ne to $^{32}$Ne.  
Dytrych $et$ $al$. use symmetry-adapted no-core shell model (SA-NCSM) in Ref. \cite{Ne7} to study the shape of $^{20}$Ne g.s. and other excited states. The authors have used Sp(3, $R$)-adapted basis for this work which reduces the size of the Hamiltonian matrix significantly, and a single symplectic basis provides information about nuclear dynamics. Recently, a machine learning (ML) technique based on neural network  was developed using $ab$ $initio$ SA-NCSM results of some selected light nuclei as training data \cite{Ne5*}. In the Ref. \cite{charge_radii2}, $ab$ $inito$ no-core Monte Carlo shell model approach is used to calculate the ground-state energies and point-proton root-mean-square radii of light $4N$ self-conjugate nuclei up to $^{20}$Ne. For this work, the authors employed two nonlocal $NN$ interactions, namely JISP16 and Daejeon16. Stroberg $et$ $al.$ \cite{Ne5**} use the valence-space in-medium similarity renormalization group (VS-IMSRG)  to evaluate the isospin dependence of the underprediction of $E2$ strength. For this purpose, different $sd$-shell nuclei, including $^{18,19,21,22}$Ne isotopes, are used and they find out that the missing strength is mostly isoscalar in nature. Another approach, namely the energy density functional (EDF) framework is used to study the even-neon isotopes starting from $^{20}$Ne towards the {\color{black} drip line} nucleus in Ref. \cite{Ne6}. The authors have investigated the evolution of cluster structure in the even-even neon isotopes towards neutron drip line \cite{dripline1}.  
Experimentally the neutron drip lines for neon and fluorine have been observed at the RIKEN Radioactive Isotope Beam Factory by fragmentation of an intense beam of $^{48}$Ca on a thick Be target \cite{dripline2}. This work indicates that $^{34}$Ne is the heaviest bound isotope of the neon isotopic chain. 


The charge radius of an atomic nucleus is another fundamental property that is helpful in understanding the nuclear structure. Due to the complex nature of nuclear interaction, the charge radius reflects several nuclear structure phenomena such as halo structures,
the occurrence of magic numbers, 
shape staggering, 
shape coexistence, 
pairing correlations etc. \cite{charge.rad1}. Therefore, experimental measurements of charge radii along with theoretical validations and predictions are essential to understand nuclear structure properties. Different experimental methods are available to measure the charge radii of nuclei across the nuclear chart. Some methods like the e$^-$ scattering experiments provide information about the charge radii directly. In contrast, other methods like the optical isotope shifts give information about isotopic changes of charge radii with respect to a stable isotope. These two types of experiments are complementary, and they can provide valuable information for charge radii of nuclei \cite{charge.rad2}. 
In Ref. \cite{Ne10}, high-precision mass and charge radius of $^{17-22}$Ne are measured using Penning trap mass spectroscopy and collinear laser spectroscopy. This paper mainly discusses the charge radius of two-proton-halo candidate $^{17}$Ne, which shows an unusually large charge radius than the other members of that isotopic chain. 
 Marinova, $et.$ $al.$ in Ref. \cite{Ne9},  measured the changes in mean-square charge radii of neon isotopes from $^{17}$Ne up to $^{28}$Ne with respect to the stable $^{20}$Ne isotope using laser spectroscopy measurements.   
In Ref. \cite{Ne8}, charge radii of neon and magnesium isotopes are calculated along with the ground-state energies and two neutron separation energies using $ab$ $initio$ coupled-cluster (CC) theory. 
In this work, the authors have used $NN$ and $3N$ potentials mainly based on $\Delta$NNLO$_{GO}$ interaction.
Like the charge radii, determining point proton radii of neutron-rich nuclei provides valuable information about the underlying structure. The point proton radii of $^{17-22}$N were reported for the first time in Ref. \cite{N=14} by measuring the charge changing cross-sections ($\sigma_{cc}$). The $r_p$ measured from $\sigma_{cc}$ are compared with the direct measurement of $r_p$ from e$^-$ scattering and theoretical predictions from the shell model and $ab$ $initio$ approaches: VS-IMSRG and {\color{black}coupled}-cluster (CC) method. All this information shows a local minimum of $r_p$ at $^{21}$N and the evolution of a halo-like structure of $^{22}$N. These two observations hint  a shell gap at $N$ = 14 in the nitrogen isotopic chain.

In the present work, our aim is to study $^{18-24}$Ne isotopes within the framework of NCSM using INOY \cite{INOY}, CDB2K \cite{CDBonn} and chiral N$^3$LO \cite{N3LO}  $NN$ interactions. For $^{18}$Ne, we have reached basis size up to $N_{max}$ = 6 and for the rest of the neon isotopes basis size up to $N_{max}$ = 4 have been reached. We have reached  the highest m-scheme dimension 1.0 $\times$ 10$^{9}$ for $N_{max}$ = 4 in the case of $^{24}$Ne. In addition to the low-lying energy spectra of these neon isotopes, we have also calculated {\color{black} electromagnetic} properties and point-proton radii. 


This paper is organized as follows: In Sec. \ref{sect2}, we briefly describe the basic formalism of the NCSM approach. Different $NN$ interactions used in this paper are described briefly in Sec. \ref{sect3}. In Sec. \ref{sect4}, the NCSM results for energy spectra are presented. The NCSM results of the g.s. energies and  electromagnetic properties 
are discussed in Sec. \ref{sect5}.
The point-proton radii results are reported in  Sec. \ref{sect6}. Finally, we have summarized our work in Sec. \ref{sect7}.

\section{No-core shell model formalism}
\label{sect2}
The no-core-shell model (NCSM) \cite{NCSM_r1, NCSM_r2} is an $ab$ $initio$ approach which is very successful in describing nuclear properties in the lighter mass region. It is a non-perturbative approach in which a system of non-relativistic nucleons  is considered that interact via realistic $NN$ or $NN$+$3N$ interactions among themselves. Unlike the conventional shell model, there is no core in this approach, and all the nucleons are considered active.\

The NCSM Hamiltonian considering only realistic $NN$ interactions among the nucleons, is given by
\begin{equation}
\label{eq:(1)}
H_{A}= T_{rel} + V = \frac{1}{A} \sum_{i< j}^{A} \frac{({\vec p_i - \vec p_j})^2}{2m}
+  \sum_{i<j}^A V^{NN}_{ij} 
\end{equation}
where, $T_{rel}$ denotes the relative kinetic energy, $p_{i,j}$ (\textit{i,j}= 1, 2, 3,..., A) is momentum of a single nucleon, $m$ is the mass of nucleon and $V^{NN}_{ij}$ is the \textit{NN} interaction which contains the nuclear part along with the Coulomb part.

In the NCSM, the eigenstates of the Hamiltonian are determined by solving a large-scale matrix eigenvalue problem within a model space spanned by a set of many-body states. The model space of the NCSM is constructed by taking Slater determinants of the harmonic-oscillator single-particle states having $N_{max}$-truncation on the many-body states.  
The truncation parameter $N_{max}$ measures the allowed harmonic-oscillator quanta above the unperturbed g.s. of an A-nucleon system. The use of harmonic-oscillator basis along with $N_{max}\hbar\Omega$ truncation allows the separation of the intrinsic component of the many-body basis from the center-of-mass component for all $N_{max}$. In order to treat the original Hamiltonian (\ref{eq:(1)}) in a truncated HO basis, it is necessary to derive the effective interaction suitable for the basis truncation.
Some renormalization techniques based on similarity transformation are necessary to obtain an effective interaction. Two such techniques that are utilized in the NCSM are: the Okubo-Lee-Suzuki (OLS) \cite{OLS1, OLS2, OLS3, OLS4} scheme and the similarity renormalization group (SRG) \cite{SRG1, SRG2}. 
We have used the OLS scheme to derive effective interactions in this work. 

The derivation of the effective interaction involves the modification of the original Hamiltonian (\ref{eq:(1)}) by adding to it the center-of-mass Hamiltonian ($H_{cm}$):

\begin{equation}
	\label{eq:(2)}
	H^{\Omega}_A = H_A + H_{CM} = \sum_{i = 1}^A [\frac{p_i^2}{2m} + \frac{1}{2}m \Omega^2 r_i^2]
	+ \sum_{i < j}^A [V^{NN}_{ij} - \frac{m \Omega^2}{2 A} (r_i - r_j)^2]
\end{equation}

where, $H_{CM}$ = $T_{CM}$ + $U_{CM}$. As $H_A$ is a translational invariant Hamiltonian, the addition of $H_{CM}$ to it does not change the fundamental property of the system, and it is subtracted out in a subsequent step. 
The effective Hamiltonian contains all the terms up to A-body, even if the original Hamiltonian consist of only two-body terms. As we have considered only the two-body part of the Hamiltonian, we expect that the two-body part of the effective interaction will be the dominant part of the exact effective interaction. So, in this work, only the two-body OLS effective interactions are taken into account, and it is obtained by solving the Hamiltonian (\ref{eq:(2)}) exactly for a two nucleon system.  

In the final step, the c.m. Hamiltonian $H_{c.m}$ is subtracted out from Hamiltonian (\ref{eq:(2)}) and the Lawson projection term \cite{Lawson} $\beta (H_{c.m} - \frac{3}{2} \hbar \Omega)$ is added to it. In this work, the value of $\beta$ is taken to be 10. 
The addition of the  Lawson projection term does not shift the physical energy states having a passive state of the c.m. motion. However, it does shift the states with excited c.m. motion to a higher value, so that there is no mixing of the physically relevant states and the states with excited c.m. motion. In the OLS scheme, the infinite A-nucleon HO basis is divided into the finite active space ($P$) consisting of all states up to $N_{max}$ HO excitations and the excluded space ($Q = 1 - P$). The effective Hamiltonian for the A-nucleon system, considering only the two-body effective interaction, is given by:

\begin{equation}\label{eq:(3)}
\begin{split}
&\hspace*{-1.2cm} H_{A, \textit{eff}}^{\Omega} = P \Bigg\{ \sum_{i<j}^{A} \bigg[ \frac{{(\vec{p}_i - \vec{p}_j)}^2}{2mA} + \frac{m {\Omega}^2}{2A} {(\vec{r}_i - \vec{r}_j)}^2 \bigg] \\
& + \sum_{i<j}^{A} \bigg[ V^{\rm NN}_{ij} - \frac{m {\Omega}^2}{2A}{(\vec{r}_i - \vec{r}_j)}^2 \bigg]_{\text{eff.}}
\hspace{-0.3cm} 
+ \beta \bigg( H_{c.m.} - \frac{3}{2}\hbar\Omega \bigg) \Bigg\} P \ 
\end{split}
\end{equation}

The effective interaction of Eq. (\ref{eq:(3)}) depends on nucleon number A, the HO frequency $\Omega$ and the basis size of the P-space controlled by $N_{max}$. In order to minimize the effect of neglected higher clusters in the derivation of effective Hamiltonian, large model space is needed. For the case of $N_{max}$ $\to$ $\infty$, the NCSM effective Hamiltonian given above approaches the bare Hamiltonian (\ref{eq:(1)}). Accordingly, the NCSM results converge to the exact solution as the basis size increases.


\section{\textit{NN} interactions}
\label{sect3}
In the present work, we have used three realistic $NN$ interactions namely, inside nonlocal outside Yukawa (INOY) \cite{INOY, INOY2, INOY3}, charge dependent Bonn $NN$ interaction (CDB2K) \cite{CDB2K3, CDBonn} and N$^3$LO \cite{QCD, N3LO}. 


The inside nonlocal outside Yukawa (INOY) \cite{INOY, INOY2, INOY3} is a phenomenological potential with a nonlocal part at a short distance ($\le$ 1 fm). At a large distance ($\ge$ 3 fm), it becomes the Yukawa tail of local Argonne $v18$ \cite{av18} potential. The nonlocality of nuclear interactions at short distances is primarily due to the internal structure of the nucleon. There is a smooth {\color{black} cut-off} in the transition region (1-3 fm) from the local to the nonlocal part, and the range of locality and nonlocality can be controlled explicitly. This $NN$ interaction is defined as
\begin{equation}
V^{full}_{ll'}(r,r')  = W_{ll'}(r,r') + \delta (r-r') F^{cut}_{ll'}(r) V^{Yukawa}_{ll'}(r) 
\end{equation}
where, the cut-off function is 
\begin{equation}
\begin{split}
&\hspace*{-1.25cm}F^{cut}_{ll'}(r) = 1- e^{-[\alpha_{ll'}(r-R_{ll'})]^{2}}, ~\text{for} ~ r \geq R_{ll'}\\
& = 0,  \hspace*{3cm}\text{for} ~r\leq R_{ll'}
\end{split}
\end{equation}

and $W_{ll^{'}}(r,r^{'})$ and $V_{ll^{'}}^{Yukawa}(r)$ are the phenomenological nonlocal part and the local Yukawa tail of Argonne $v18$ potential, respectively. The parameters $\alpha_{ll'}$ and $R_{ll'}$ are considered to be independent of the angular momenta having values 1.0 fm$^{-1}$ and 1.0 fm, respectively. The other parameters of W$_{ll'}(r,r')$ are determined by fitting $NN$ data and the binding energy of $^3$He. This interaction model breaks charge independence and charge symmetry as it is essential to reproduce all low-energy experimental parameters, including $np$, $pp$, and $nn$ scattering lengths to high precision. To describe a few-nucleon system reasonably well, additional $3N$ forces are necessary along with local $NN$ potentials, but INOY $NN$ interaction does not need $3N$ force to provide the binding energies of $3N$ systems. The effects of three-nucleon forces are partly included in the non-local part of this interaction.

The charge-dependent Bonn 2000 (CDB2K) \cite{CDBonn} is a one-boson exchange $NN$ potential which includes all the mesons $\pi$, $\eta$, $\rho$, and $\omega$ with masses below the nucleon mass. Due to the vanishing coupling of $\eta$-meson to the nucleon, it is dropped from the potential model. In addition to these mesons, two partial-wave dependent scalar-isoscalar $\sigma$ bosons are also introduced. This charge-dependent potential reproduces accurately charge symmetry breaking (CSB) and charge independence breaking (CIB) in all partial waves with J$\le$4 as predicted by the Bonn full model \cite{CDB2K3}. In this model, three $NN$ interactions are constructed: a proton-proton, a neutron-neutron, and a neutron-proton potential, and the differences between them are measured by CSB and CIB. These potentials fit the world proton-proton data available in the year 2000 with $\chi^2$/datum = 1.01 and the corresponding neutron-proton data with $\chi^2$/datum = 1.02 below 350 MeV. The CDB2K potential uses the original form of the covariant Feynman amplitudes of meson exchange without local approximation. So, the off-shell {\color{black} behaviour} of this $NN$ potential is different from other local $NN$ potentials.

The chiral $NN$ interaction at next-to-next-to-next-to-leading order (N$^3$LO) \cite{N3LO} is derived from the quantum chromodynamics using chiral perturbation theory ($\chi$PT). The $\chi$PT is a systematic expansion of the nuclear potential in terms of (Q/$\Lambda_\chi$)$^\nu$, where Q is a low-momentum scale or pion mass, $\Lambda_\chi$ $\approx$ 1 GeV is the chiral symmetry breaking scale and $\nu$ $\geq$ 0 \cite{QCD}. For a definite order $\nu$, there is a finite number of unique and calculable contributing terms. The long-range part of the nuclear interaction is associated with the exchange of pions, whereas the short-range part is defined in terms of contact terms. The charge dependence of this interaction is important for a good fit of the low-energy $pp$ and $np$ data. There are, in total, 29 parameters of the N$^3$LO potential. Three of them are low-energy constants (LECs), $c_2$, $c_3$ and $c_4$ that appear in the $\pi N$ Lagrangians. The most important fit parameters are 24 contact terms that dominate the partial waves with L $\le$ 2, and the remaining two parameters are two charge-dependent contact terms. The $NN$ interaction at this order can be compared to high-precision phenomenological potential AV18 \cite{av18} in terms of the accuracy in reproducing the $NN$ data below 290 MeV lab-energy.

The NCSM results reported in this paper have been calculated using the pAntoine \cite{pAntoine1, pAntoine2, pAntoine3} code. 
Our group has previously reported the NCSM results for N \cite{arch1}, C \cite {pr1,pr2}, O \cite {arch2}, F \cite{arch2} and B \cite{priyanka}  isotopes.

\begin{table*}[ht]
	\caption{
	The m-scheme dimensions of Ne isotopes corresponding to different $N_{max}$ are shown. The dimensions up to which we have reached are shown in boldface.}

\label{tab:my_label}
	\hspace{-1cm}
		\begin{center}
		
		\begin{tabular}{cccccccccccccc}
			\hline
			\hline  
			\vspace{0.75mm}
			$N_{max}$  &  $^{18}$Ne   &  $^{19}$Ne   &  $^{20}$Ne    & $^{21}$Ne   & $^{22}$Ne    & $^{23}$Ne   & $^{24}$Ne \\
			\hline \vspace{-2.9mm}\\
			0 & \textbf{14}     & \textbf{128} & \textbf{ 640} & \textbf{ 1935} & \textbf{ 4206}  & \textbf{ 6457} & \textbf{ 7562} \\
			2 & \textbf{2.7 $\times$ 10$^{4}$ }  &\textbf{1.4 $\times $10$^{5}$ }  &\textbf{5.4 $\times$ 10$^{5}$} & \textbf{1.5 $\times$ 10$^{6}$ } & \textbf{3.1 $\times$ 10$^6$  }  & \textbf{5.0 $\times$ 10$^{6}$ } & \textbf{6.5 $\times$ 10$^{6}$ } \\
			4 & \textbf{5.1 $\times$ 10$^{6}$ }  & \textbf{2.2 $\times$ 10$^{7}$ }  & \textbf{7.5 $\times$ 10$^{7}$ } & \textbf{1.9 $\times$ 10$^{8}$ }  & \textbf{4.1 $\times$ 10$^{8}$}  & \textbf{7.1 $\times$ 10$^{8}$ } & \textbf{1.0 $\times$ 10$^{9}$}\\
			6 & \textbf{3.4 $\times$ 10$^{8}$}  & $1.3\times10^{9}$ &  $4.4 \times10^{9}$ &  $ 1.1 \times10^{10}$ &  $ 2.5 \times10^{10}$ &   $ 4.5\times10^{10}$ & $ 6.8\times10^{10}$ \\
			8 &  $1.2 \times10^{10}$ &  $4.7\times10^{10}$   &  $1.5\times10^{11}$  &   $ 4.0 \times10^{11} $   &   $ 9.0 \times 10^{11} $ &  $ 1.7 \times 10^{12} $ &  $ 2.7 \times 10^{12} $\\
			\hline \hline
		\end{tabular} 
		\end{center}
\end{table*}

\begin{figure*}
	 \includegraphics[width=8.2cm]{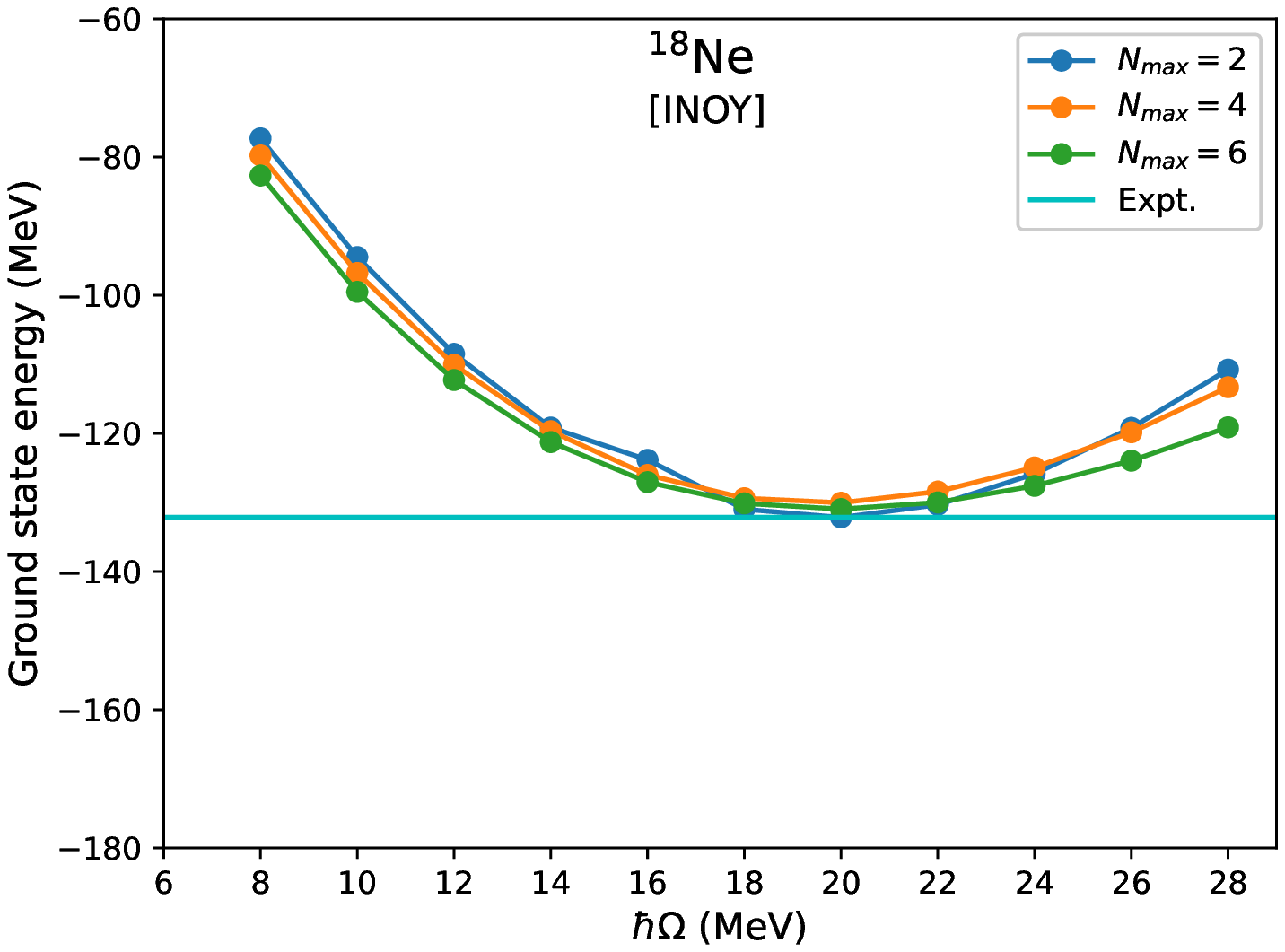}
	\includegraphics[width=8.2cm]{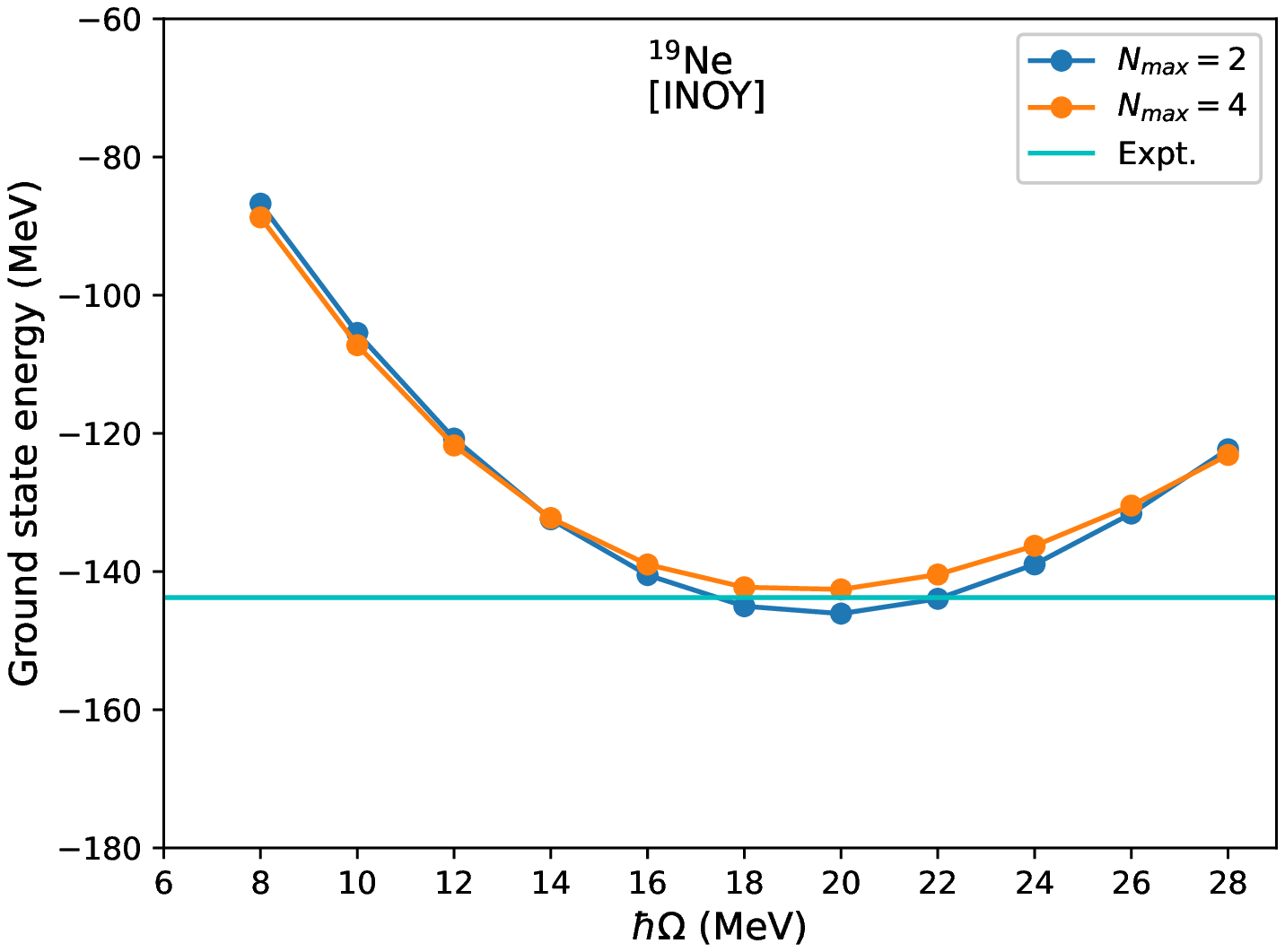}
	\includegraphics[width=8.2cm]{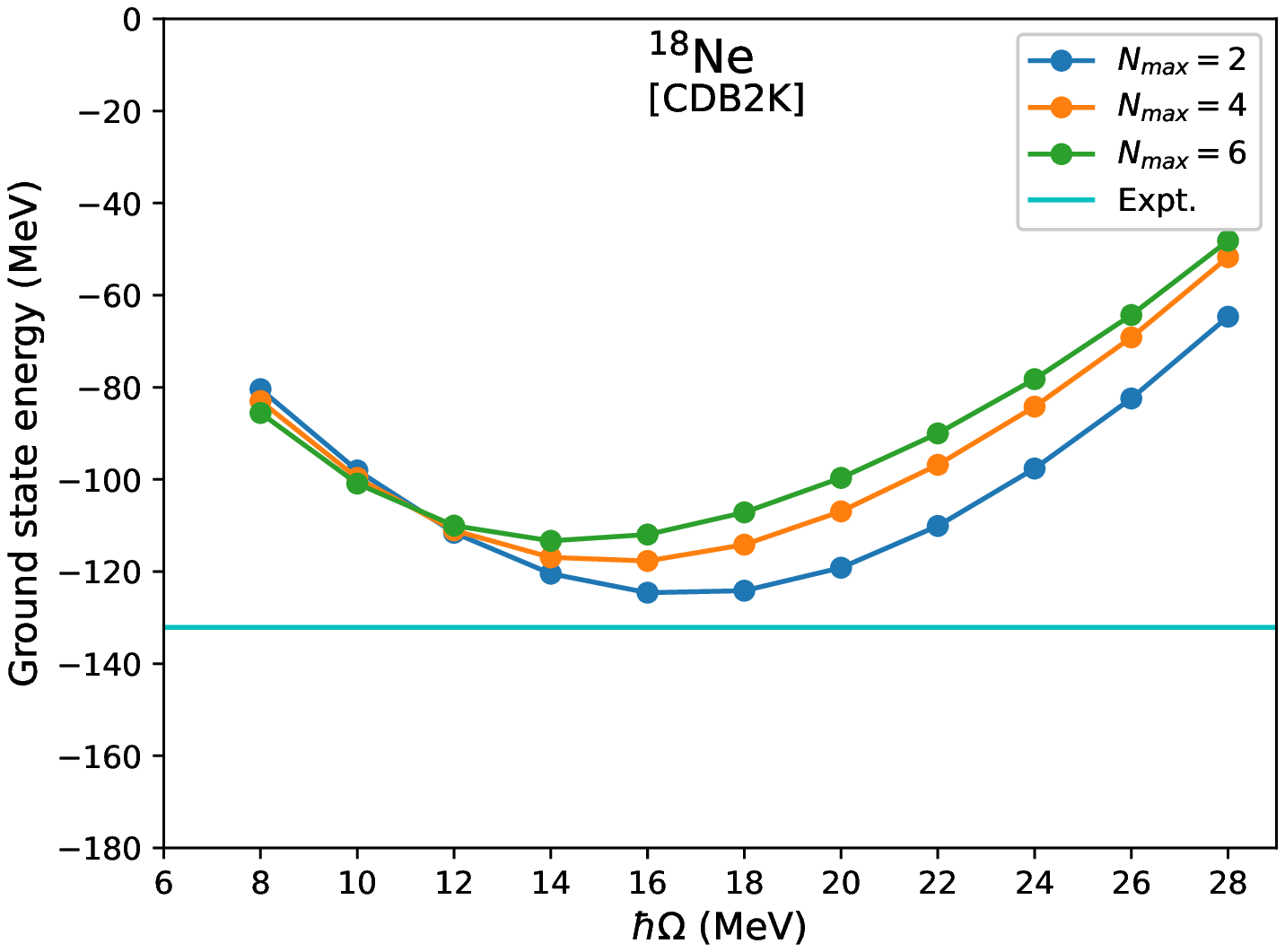}
	\includegraphics[width=8.2cm]{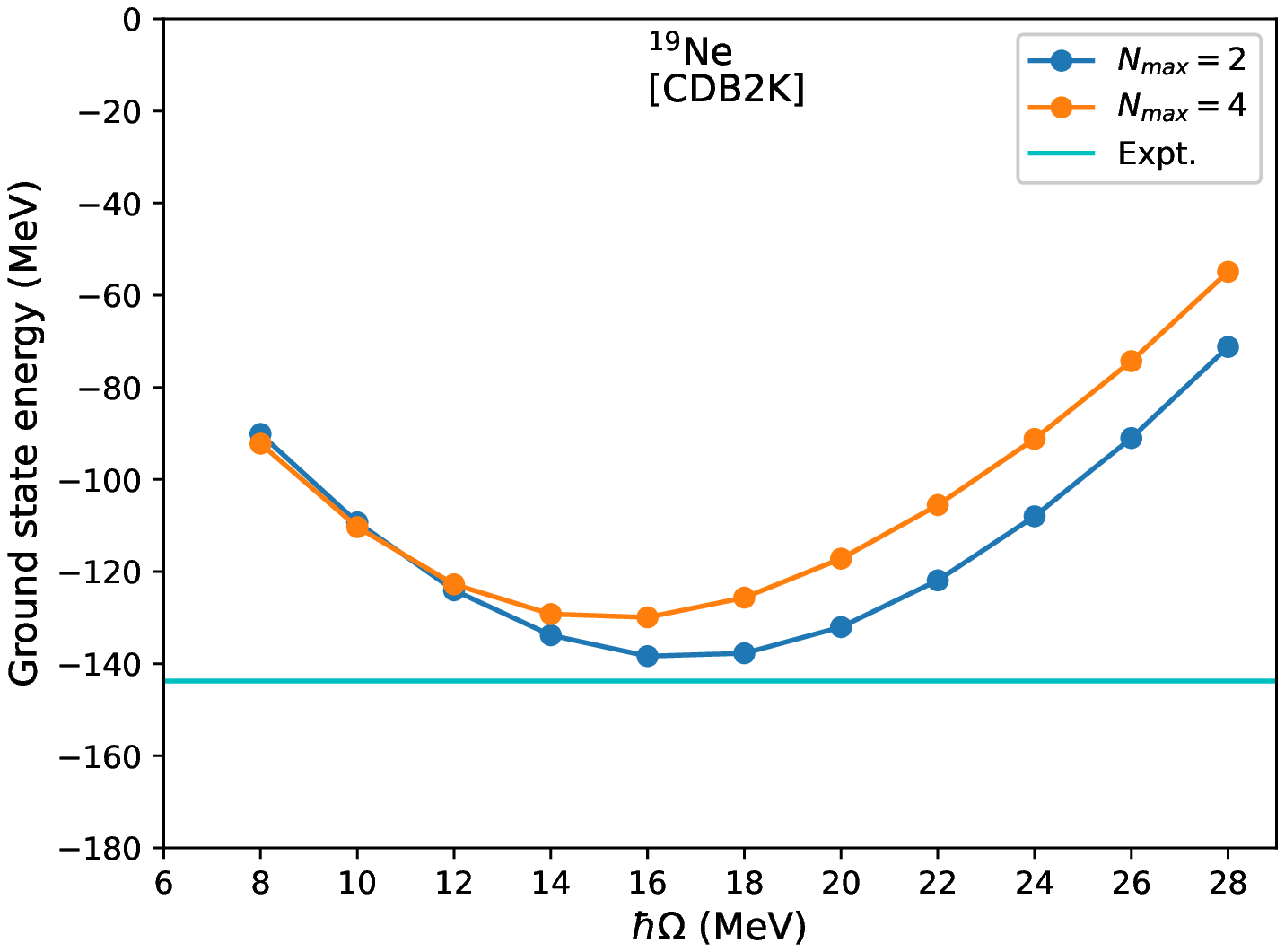}
	\includegraphics[width=8.2cm]{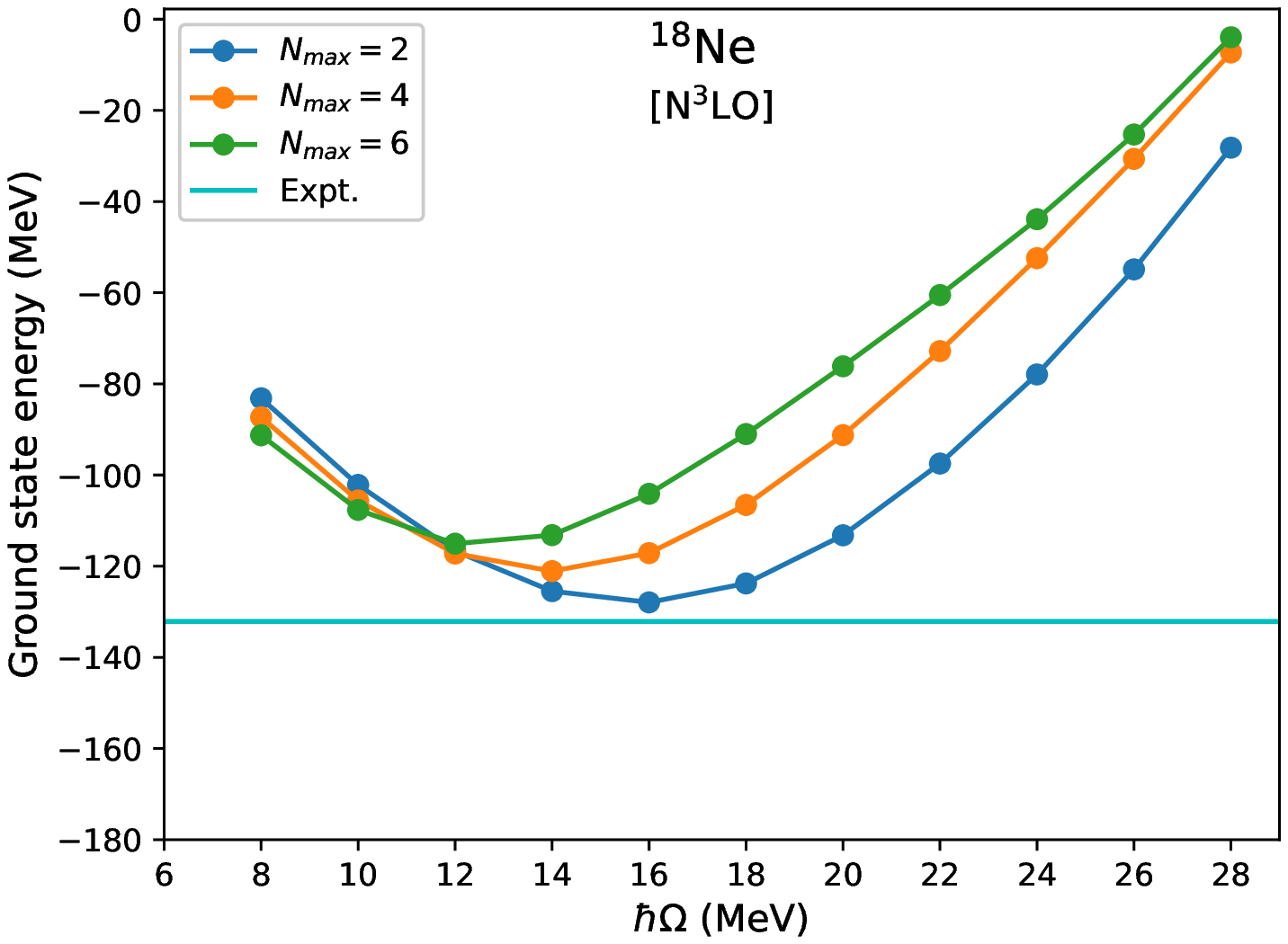}
	\includegraphics[width=8.2cm]{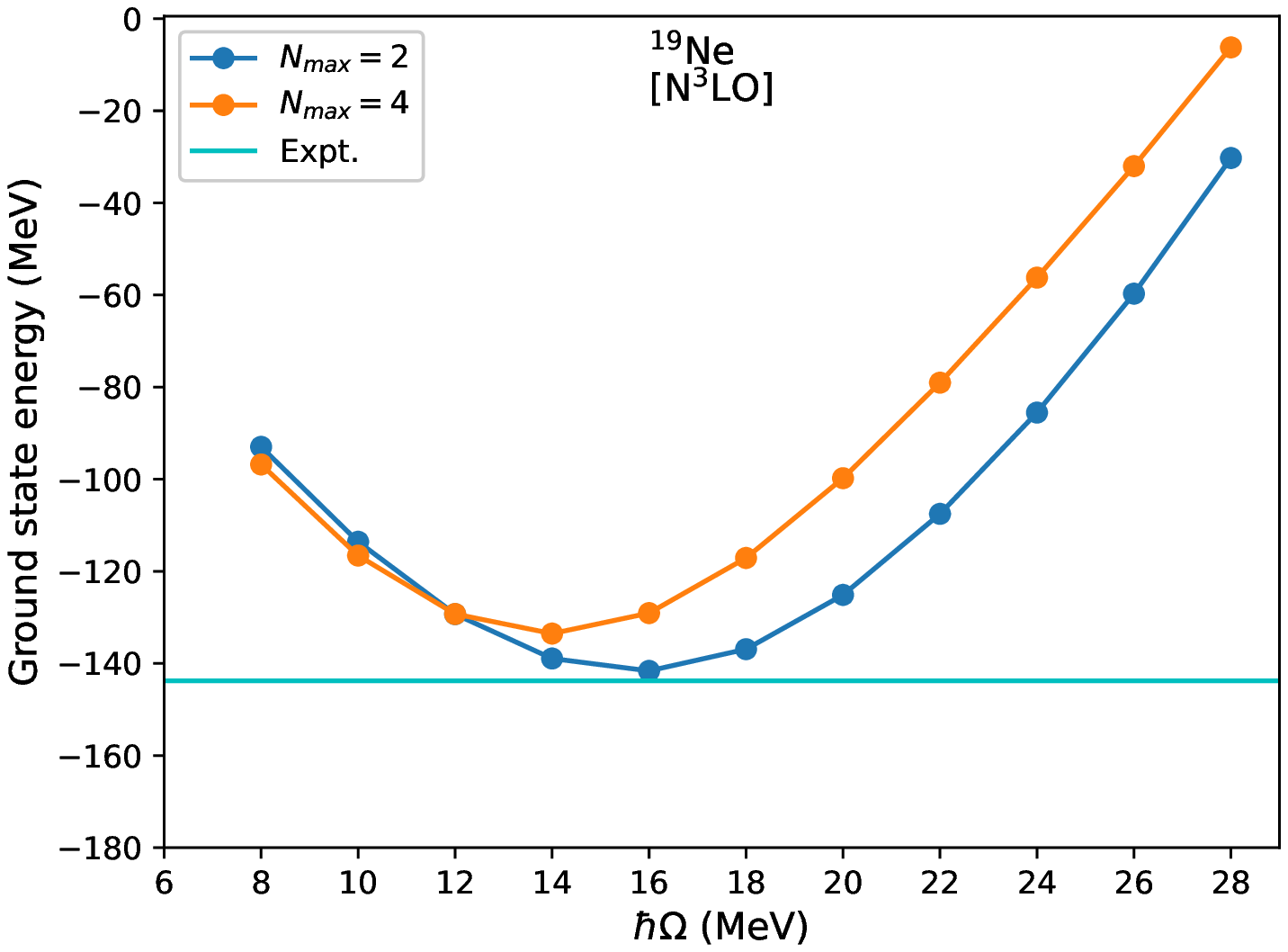}
	\caption{Variation of the g.s. energies of $^{18}$Ne and $^{19}$Ne with HO frequency for different $N_{max}$.}\label{gs_energy1}
\end{figure*}

\begin{figure*}
	\begin{center}
	\includegraphics[width=17cm, height=7.6cm]{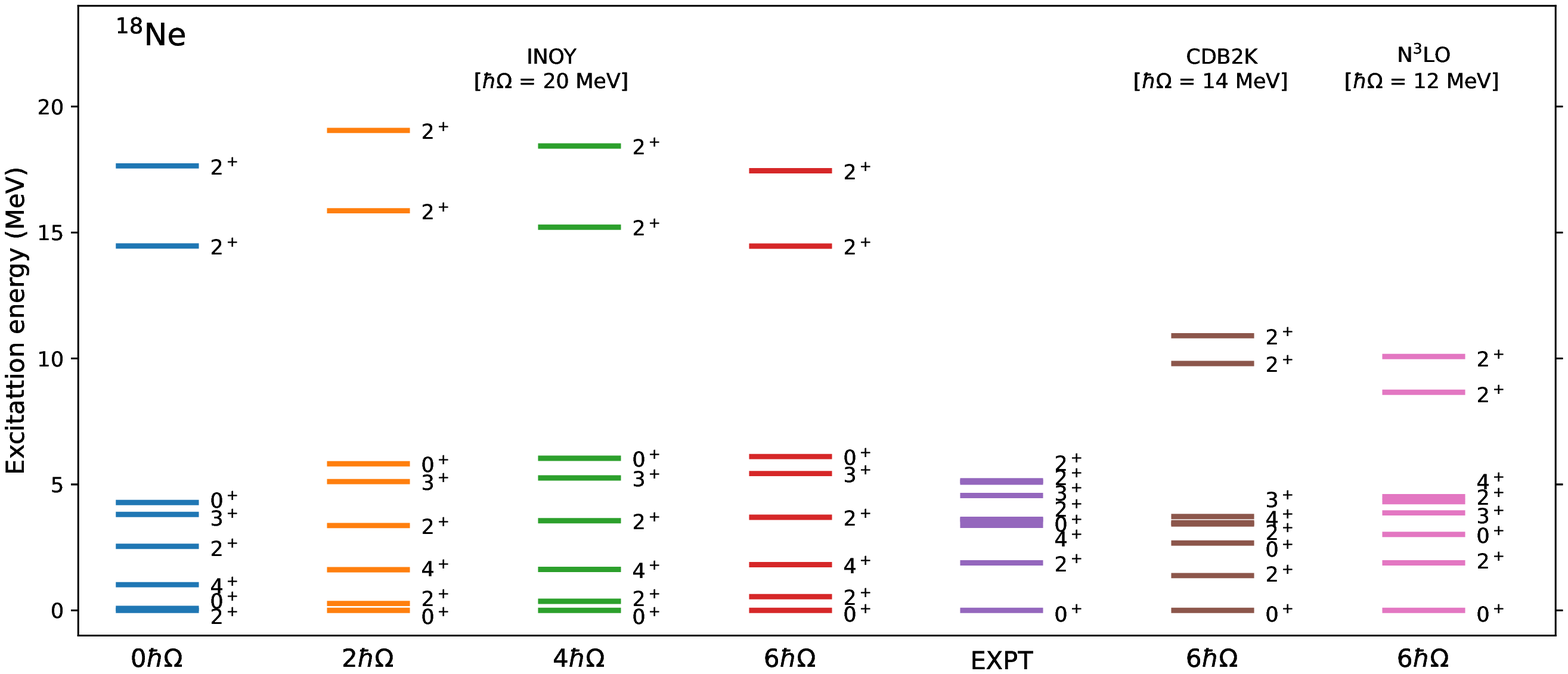}
	\includegraphics[width=17cm, height=7.6cm]{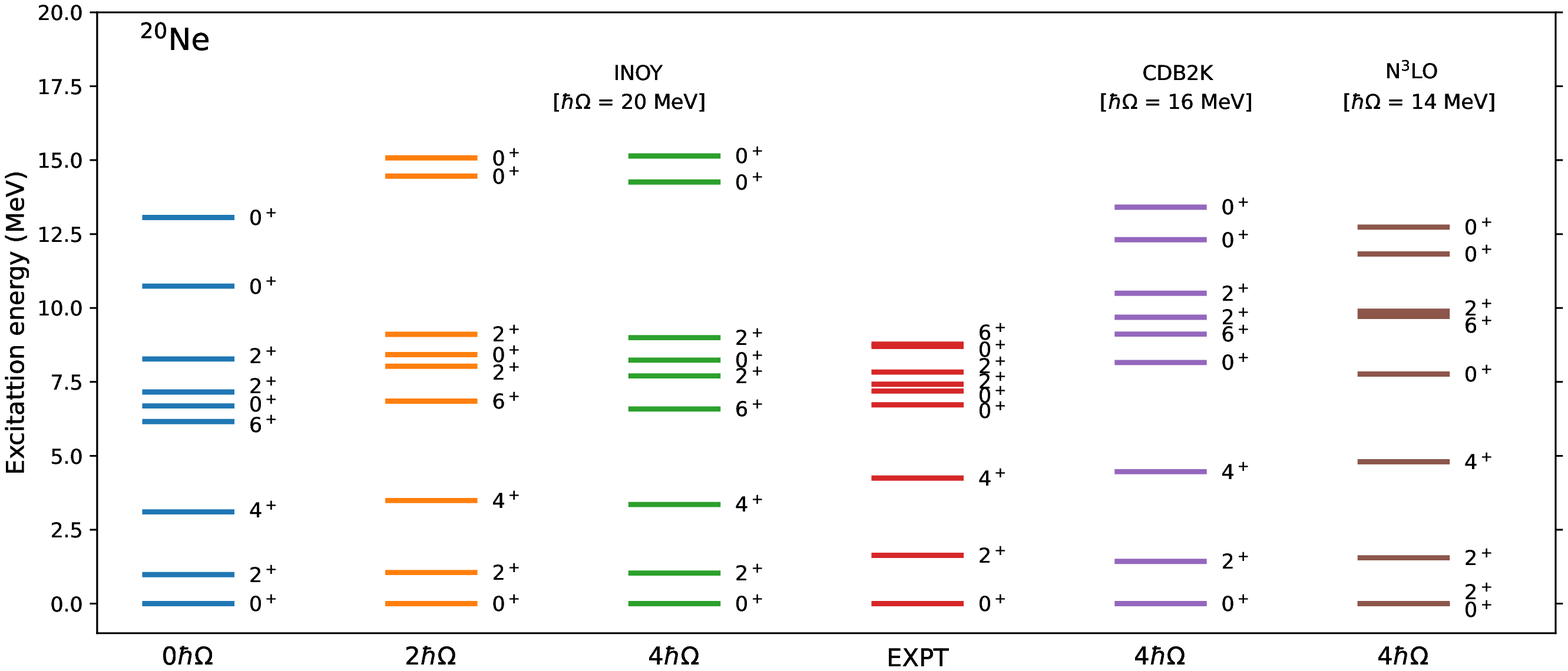}
	\includegraphics[width=17cm, height=7.6cm]{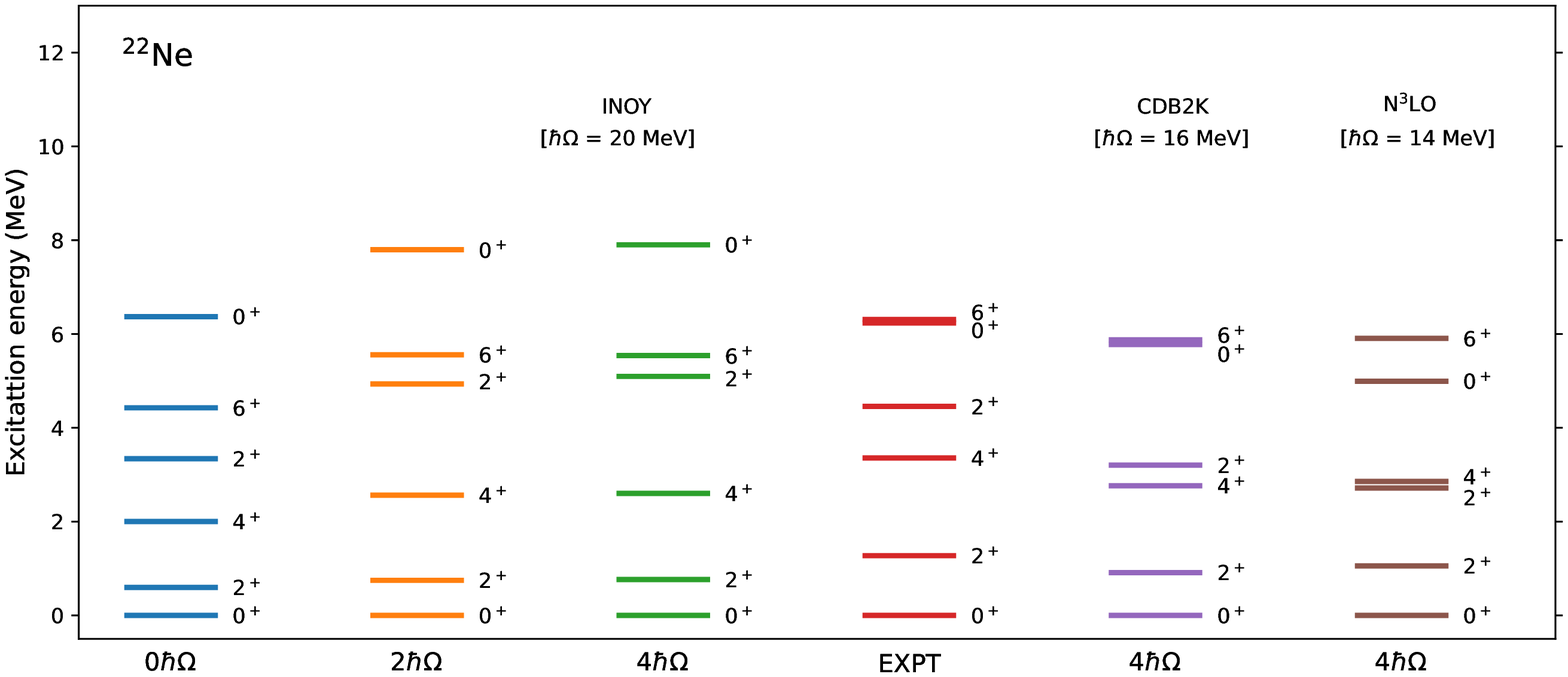}
	\caption{No core shell model results for low-lying states of $^{18,20,22}$Ne isotopes using INOY, CDB2K and N$^3$LO interactions.}\label{Even_Ne_spectra}
	\end{center}
\end{figure*}

\begin{figure*}
\begin{center}
	\includegraphics[width=17cm, height=8cm]{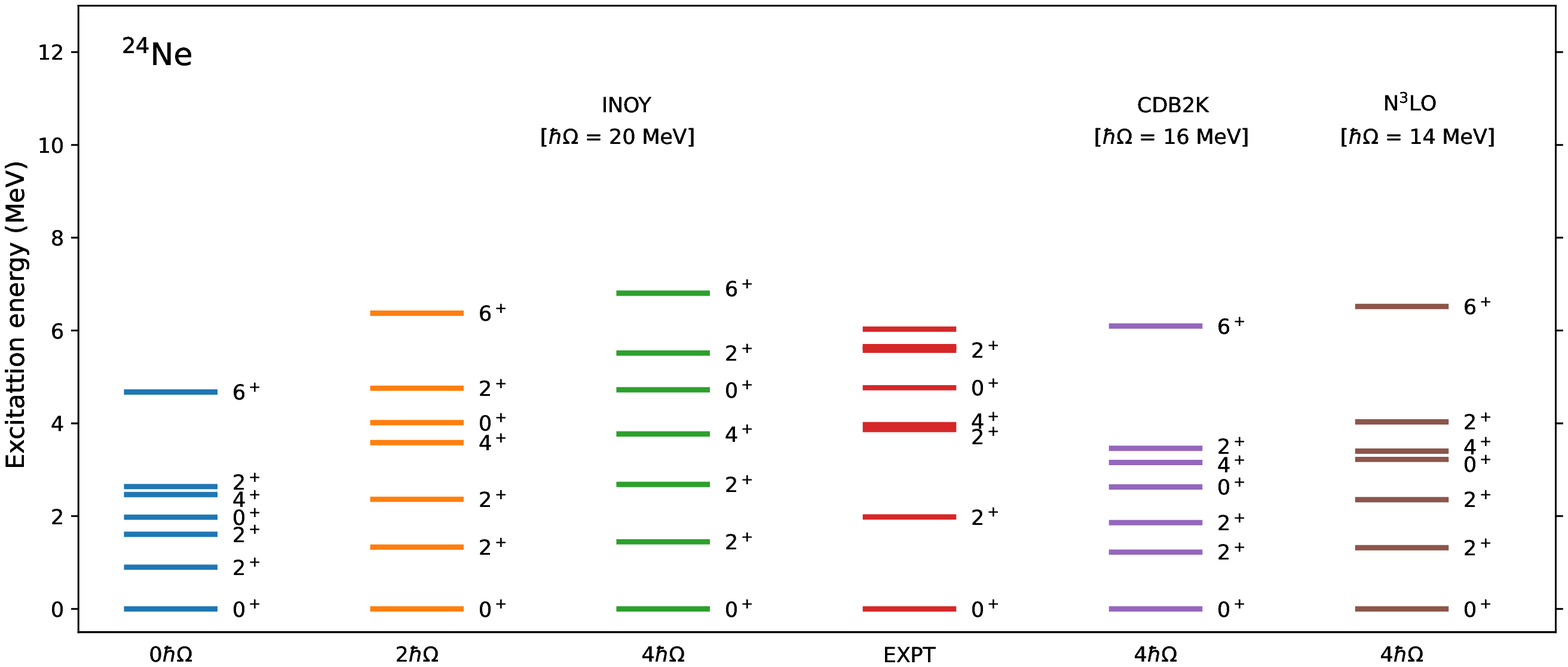}
	\caption{No core shell model results for low-lying states of $^{24}$Ne using INOY, CDB2K and N$^3$LO interactions.}\label{24Ne_spectra}
	\end{center}
\end{figure*}

\section{Results and Discussions}
\label{sect4}

In this section, we have presented the NCSM results of neon isotopes. The m-scheme dimensions for $^{18-24}$Ne isotopes corresponding to different $N_{max}$ are shown in Table \ref{tab:my_label}. The table shows that dimension of Hamiltonian matrix of Ne isotopes increases rapidly with the increase in $N_{max}$ and the number of nucleons, A. We are able to reach basis size up to $N_{max}$ = 6 for $^{18}$Ne and $N_{max}$ = 4 for  the remaining neon isotopes. The first step in the NCSM calculations is to obtain the optimum frequency for the g.s. energy, i.e., that frequency for which the g.s. energy is minimum for the maximum possible $N_{max}$. To obtain the optimum frequency for a particular interaction, the calculated g.s. energies for a particular nucleus are plotted with HO frequencies for different $N_{max}$. Fig. \ref{gs_energy1} shows the variation of g.s. energies of $^{18}$Ne and $^{19}$Ne with frequency for different $N_{max}$. For $^{18}$Ne, we show the variation from $N_{max}$= 2 to 6; while for $^{19}$Ne the variation from $N_{max}$= 2 to 4 are shown. 
From Fig. \ref{gs_energy1}, we obtain the optimal frequencies for INOY, CDB2K and N$^3$LO  for $^{18}$Ne as 20, 14 and 12 MeV, respectively corresponding to highest $N_{max}$. 
Similarly, we have obtained the optimal frequencies for other neon isotopes corresponding to different $NN$ interactions. 
The g.s. energies obtained from OLS renormalized interactions are non-variational with $\hbar \Omega$ and $N_{max}$ \cite{NCSM_r2}. 

\begin{figure*}
	\begin{center}
	\includegraphics[width=17cm, height=7.6cm]{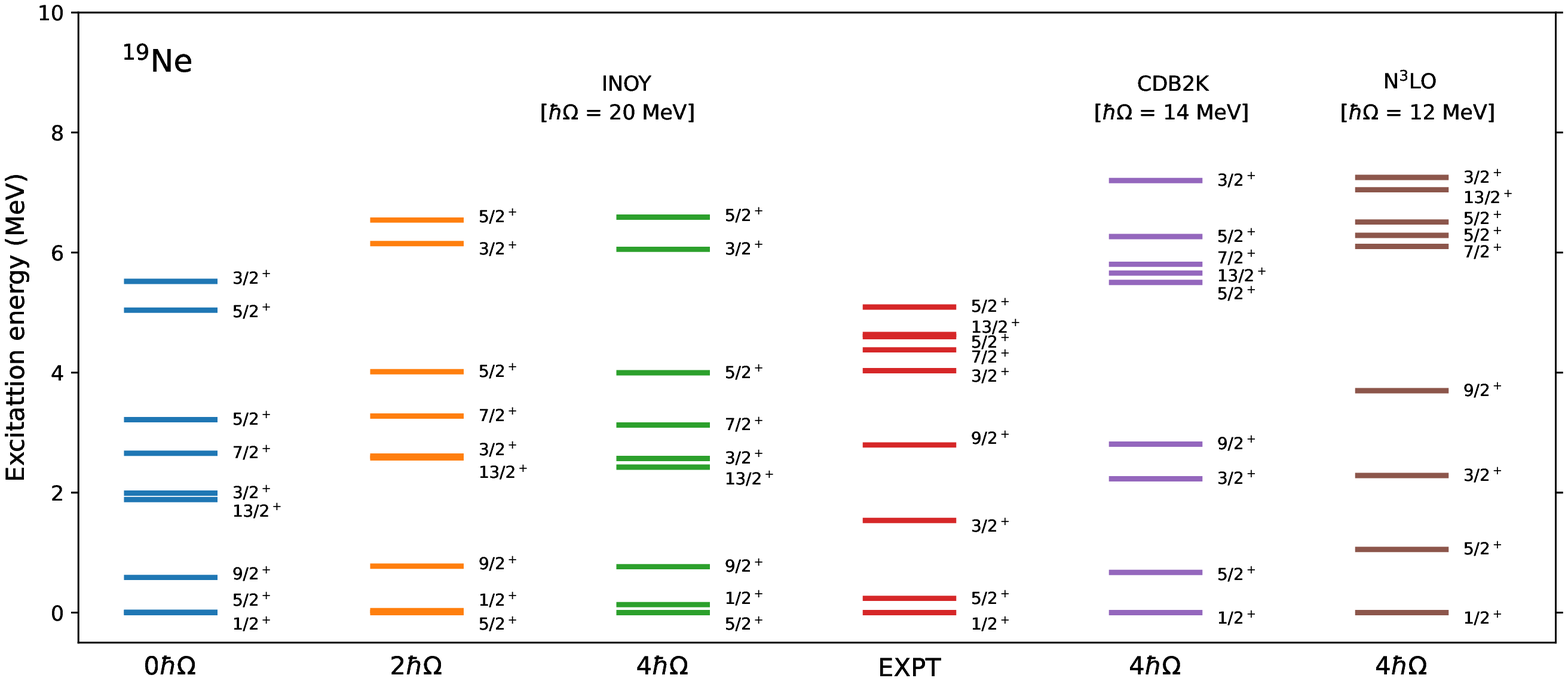}
	\includegraphics[width=17cm, height=7.6cm]{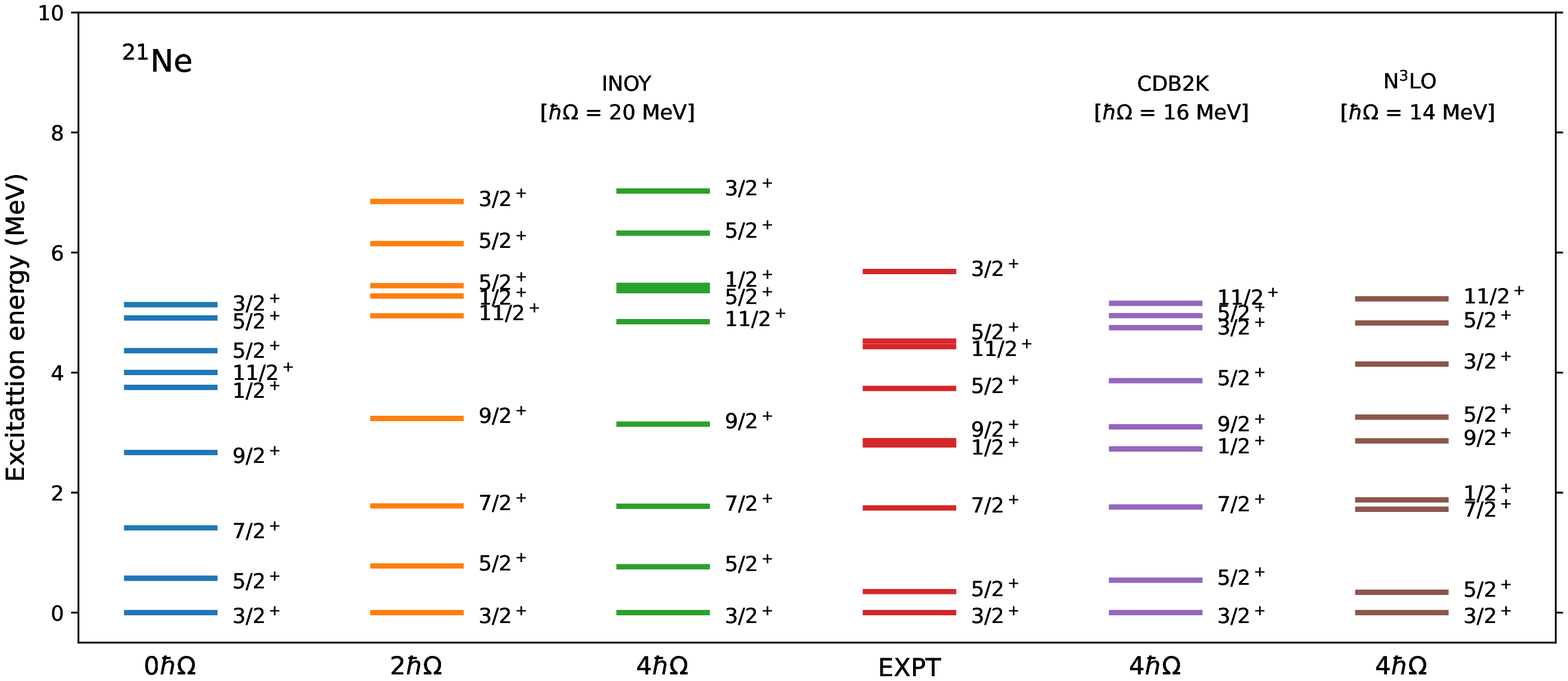}
	\includegraphics[width=17cm, height=7.6cm]{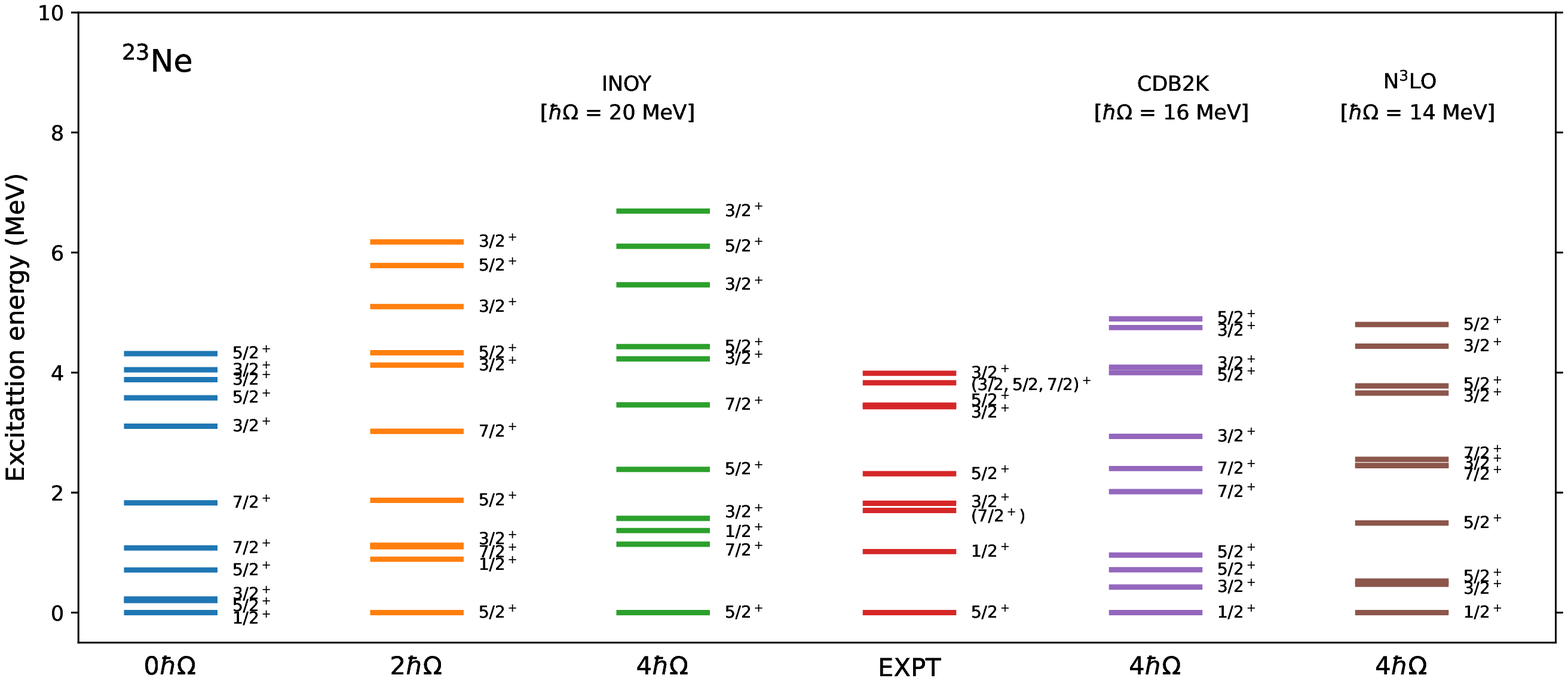}
	\caption{No core shell model results for low-lying states of $^{19,21,23}$Ne isotopes using INOY, CDB2K and N$^3$LO interactions.}\label{19Ne_spectra}	
	\end{center}
\end{figure*}

In Figs. \ref{Even_Ne_spectra}-\ref{19Ne_spectra}, we have shown the low-lying spectra of different neon isotopes. For INOY interaction, we have shown spectra from 0 to the highest $N_{max}$, and for other realistic interactions, we have shown only the results for the highest $N_{max}$. These results are compared with {\color{black} the} available experimental data. 
The optimal frequencies of the g.s. are taken to calculate low-lying states.

\subsection{Energy spectra of $^{18,20,22,24}$Ne}
For $^{18}$Ne,  the ordering of the three lowest-lying states 0$^+_1$-2$^+_1$-4$^+_1$, is well reproduced by NCSM calculations with INOY  interaction as it can be seen from the first panel of Fig. \ref{Even_Ne_spectra}. However, the separation between the g.s. and the 2$^+_1$ obtained from this interaction is small compared to the experimental value (1.887 MeV).  The excitation energy of the 2$^+_1$ state increases by 465 keV when the basis size is increased from N$_{max}$ = 0 to 6 for INOY interaction. The other two interactions, CDB2K and N$^3$LO can reproduce only the g.s. and the first excited state correctly. The excitation energy of the 2$^+_1$  obtained using N$^3$LO (1.887 MeV) matches the experimental value,  however it is lower by 500 keV with CDB2K interaction. The 4$^+_1$ state obtained for CDB2K is 102 keV higher than the experimental 4$^+_1$ state, whereas it is  1.1 MeV higher for the N$^3$LO {\color{black} interaction}. The energy separation between 4$^+_1$ and 2$^+_1$ obtained from INOY interaction decreases with basis size increment and goes close to the experimental separation (1.489 MeV). Also, the 2$^+_2$ state obtained from the INOY is near the experimental one. The calculated energy separation between 4$^+_1$ and 2$^+_2$ states for N$^3$LO is close to the experimental separation though the ordering of these two states is reversed.  Both 2$^+_3$ and 2$^+_4$ states obtained from all the three interactions are shifted upward compared to the experimental data. The 3$^+_1$ state calculated from INOY  interaction is higher than the experimental value, whereas those calculated using CDB2K and N$^3$LO are lower.

For $^{20}$Ne, the excitation energy of the first excited state, 2$^+$ is 1.633 MeV. The calculated excitation energies of this state are obtained to be 1.034 and 1.431 from INOY and CDB2K interactions, respectively. However, it is found to be very small ( $<$ 0.001 MeV) for N$^3$LO interaction. The correct ordering of 0$_1^+$-2$_1^+$-4$_1^+$-0$_2^+$ states are obtained from CDB2K interaction, but the 0$_2^+$ state is shifted upward significantly compared to the experimental 0$_2^+$ state. Also, the INOY interaction correctly reproduce spectra up to the second excited state, 4$^+_1$.  The 4$^+_1$ (6$^+_1$) states obtained from CDB2K are 216 (338) keV higher than the experimental excitation energies whereas they are 550 (939) keV higher than the experimental energies for N$^3$LO interaction.  The separation between 0$^+_3$ and 0$^+_4$ states obtained for INOY, CDB2K, and N$^3$LO are almost the same, but it is smaller than the experimental value of separation (1.509 MeV) between those two states. From the second panel of Fig. \ref{Even_Ne_spectra}, it can be seen that up to 5 MeV, the results of CDB2K are in good agreement with the experiment for $^{20}$Ne. 
 In Ref. [35], neon isotopes are studied in the framework of the projected generator coordinate method
(PGCM) and quasi-exact in-medium no-core shell model (IM-NCSM). The low-lying energy spectra {\color{black} presented for $^{20}$Ne corresponding to PGCM and IN-NCSM  are consistent with the experimental excitation energies}.

Experimentally, the excitation energy of the first excited state, 2$^+$ of $^{22}$Ne, is 1.274 MeV. The calculated energy for this state is seen to be increasing slowly for INOY interaction with the increase in the basis size from $N_{max}$ = 0 to $N_{max}$ = 4 and approaching the experimental data. Similarly, the 4$^+$ state is also slowly shifted upwards towards the experimental excitation energy level with increasing basis size.  The excitation energies of 2$^+_1$ (4$^+_1$) states increase by 168 (598) keV when the basis is increased from N$_{max}$ = 0 to 4 for INOY interaction. All the interactions except N$^3$LO reproduce the correct ordering of states: 0$_1^+$-2$_1^+$-4$_1^+$-2$_2^+$. The energy separation between the states 4$^+_1$ and 2$^+_1$ calculated by using INOY, CDB2K, and N$^3$LO are almost the same, which is slightly less than the experimental energy separation. The calculated energy separation between 6$^+_1$ and 0$^+_2$ for CDB2K is near {\color{black} to the experimental energy difference. The separation between 6$^+_1$ and 2$^+_2$ for INOY interaction is smaller than the experimental value whereas for other two interactions they are larger. }

In the case of $^{24}$Ne, the INOY interaction reproduces the correct ordering of states: 0$_1^+$-2$_1^+$-2$_2^+$-4$_1^+$-0$_2^+$-2$_3^+$,  while CDB2K and N$^3$LO reproduce correct ordering up to the second excited state, 2$^+_2$ only. 
The excited  2$_1^+$, 2$_2^+$, 4$_1^+$, 0$_2^+$, 2$_3^+$ states obtained for INOY interaction improve significantly with increasing the basis size from $N_{max}$ = 0 to 4 and approach the experimental data.  For instance, the 0$^+_2$  and 2$^+_3$ states {\color{black} improve} by more than 2 MeV on increasing basis size from N$_{max}$ = 0 to 4 for INOY interaction. Among the low-lying states, these two states are only 44 and 60 keV lower than the experimental 0$^+_2$  and 2$^+_3$ states, respectively. The excitation energies of all the states obtained for other two interactions are less than the experimental values by more {\color{black} than} 600 keV.

 From the above discussions we can conclude that among the three $NN$ interactions we considered here, INOY interaction is slightly better in reproducing the low-lying spectra of even neon {\color{black} isotopes}. Also, we noticed that several states improve significantly with the increase in basis size.


It is also  important to mention here that  Stroberg {\it et al.}  in Ref.\cite{Ragnar93}, reported ground and excited states of $^{20,22,24}$Ne isotopes using \textit{ab initio} IM-SRG approach. For these isotopes spectra are too compressed with respect to the experimental data without the inclusion of 3N forces, also a significant discrepancy in the $^{20,22}$Ne g.s. energies  are reported. The IM-SRG predicted deformed g.s. for $^{20,22}$Ne, and predicted rotational yrast states in $^{20}$Ne. It was {\color{black} demonstrated} that deformation can be captured in \textit{ab initio} framework.
Jansen {\it et al.} in Ref. \cite{Jensen94}, reported results of binding energies, excited states and $B(E2)$ transition using \textit{ab initio} coupled cluster effective interaction (CCEI) for Ne and Mg chain.
For $^{19}$Ne the spacing between g.s. ($1/2^+$) and first excited state ($5/2^+$) is well reproduced with CCEI, however USDB interaction fails to reproduce this energy gap. The CCEI interaction reproduces rotational band and $B(E2)$ transitions within the {\color{black} uncertainties} estimated from chiral effective field theory (EFT).

The high-spin structure of  $^{25}$Na and $^{22}$Ne isotope using SDPF-MU, FSU, USDB and YSOX interactions are reported in Ref. \cite{Jonathan102}.
It was concluded that for $^{22}$Ne the USDB interaction result for positive-parity states along the yrast band and $B(E2)$ were in a {\color{black} reasonable}  agreement. The YSOX predict slightly lower transition strengths in comparison to the USDB interaction.

\subsection{Energy spectra of $^{19,21,23}$Ne}
For $^{19}$Ne, 1/2$^+$ is the experimental g.s., and the excitation energy of the first excited state, 5/2$^+$ is 0.238 MeV. The CDB2K and N$^3$LO interactions reproduce the correct g.s. and first excited state. The excitation energies of this first excited state obtained from these two interactions are 0.671 and 1.053 MeV, respectively, which are greater than the experimental value. As shown in the  first panel of Fig. \ref{19Ne_spectra}, the CDB2K and N$^3$LO reproduce the correct ordering of states: 1/2$_1^+$-5/2$_1^+$-3/2$_1^+$-9/2$_1^+$.  The energy separation between 3/2$^+_1$ and 5/2$^+_1$ for N$^3$LO is 1.232 MeV which is close to the experimental energy separation (1.298 MeV). Also, it can be seen that the 9/2$^+_1$ state obtained from CDB2K is close to the experimental data. However, the INOY interaction fails to predict the correct g.s. of $^{19}$Ne.  The excitation energies of the high spin state 13/2$^+_1$ obtained from INOY  interaction is less than the experimental excitation energy. In contrast, for CDB2K and N$^3$LO, the excitation energy of this state is greater than the experimental data.

In the case of $^{21}$Ne, the ordering of the low-lying energy states 3/2$_1^+$-5/2$_1^+$-7/2$_1^+$-1/2$_1^+$-9/2$_1^+$-5/2$_1^+$- obtained from CDB2K and N$^3$LO interactions match with the experimental spectra. However, for INOY  interaction, the ordering is correct only up to the second excited state.  The {\color{black} excitation} energy of the first excited state obtained for N$^3$LO interaction is 0.339 MeV which is less than the experimental energy by only 11 keV.
The excitation energies of this state are: 0.764 and 0.540  MeV for INOY and CDB2K, respectively. For INOY, the 7/2$_1^+$ state shows a significant improvement by 360 keV as we increase the basis size from $N_{max}$ = 0 to 4 and it is close to the experimental 7/2$_1^+$ state.  
The calculated energies of this state for CDB2K and N$^3$LO interactions are also close to the experimental data and the differences are only of a few keV's. Similarly, the excitation energy of the 9/2$^+_1$ state obtained from N$^3$LO is only 4 keV lower than the experimental 9/2$^+_1$ state.

For $^{23}$Ne, 5/2$^+$ is the experimental g.s., and only the INOY interaction correctly reproduce the g.s. while the  other two interactions fail to reproduce the same. None of these three realistic interactions can reproduce the correct ordering of low-energy spectra.  In the case of INOY interaction, the ordering of the first and second excited states are reversed compared to the experimental data.  While the excitation energy of the 1/2$^+_1$ state obtained from INOY interaction is 351 keV higher than the experimental data, the {\color{black} excitation} energies of 7/2$^+_1$ (3/2$^+_1$) states for the same interaction are 561 (251) keV lower than the experimental energies of these two states.
 Also, the 5/2$^+_2$ state obtained from INOY is only 72 keV higher than the experimental 5/2$^+_2$ state.
The 3/2$^+_1$ state for INOY shows a significant improvement of about 860 keV on increasing basis size from N$_{max}$ = 0 to 4. Our result for this state is expected to be much closer to the experimental data once we increase the basis size.

 From the above discussions on the odd neon {\color{black} isotopes}, we can say that the CDB2K and N$^3$LO interactions better reproduces the low-lying spectra except for the case of $^{23}$Ne. Also, to reproduce the g.s. correctly, 3N interaction may be needed for $^{23}$Ne.

\begin{figure}[h]
\begin{center}
	\includegraphics[width=9cm]{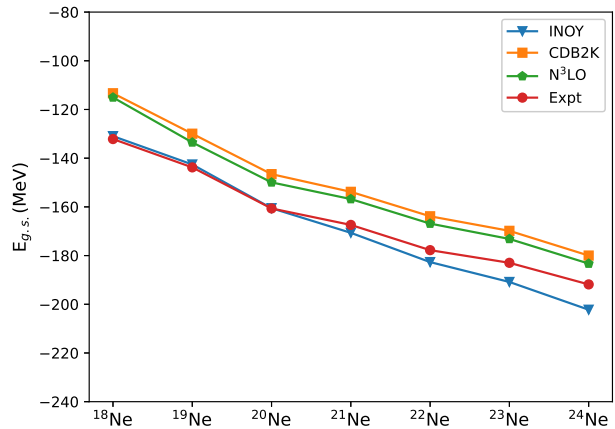}
	\caption{A-dependence of the calculated g.s. energies using  INOY, CDB2K, and N$^3$LO $NN$ interactions are compared to the experimental results. }\label{g.s.}
	\end{center}
\end{figure}

\section{Ground state energies and electromagnetic properties}
\label{sect5}

In this section, the NCSM results for the g.s. energies and electromagnetic properties are discussed.
In Table \ref{em}, we summarise the calculated results of the g.s. energies (E$_{g.s.}$), reduced electric quadrupole transition probabilities $B(E2)$ and magnetic dipole transition probabilities $B(M1)$, quadrupole (Q) and magnetic moments ($\mu$) for neon isotopes. The electromagnetic properties of the first excited states are shown in the table for even neon isotopes along with $^{19}$Ne. For $^{21, 23}$Ne, g.s. electromagnetic properties are shown. Available experimental results from Refs. \cite{NNDC, Qandmag}  and some results of recent \textit{ab initio} studies from valence-space in-medium similarity renormalization group (VS-IMSRG), PGCM and IM-NCSM as reported in Refs. \cite{Ne5,Ne5**} are also included in this table for comparison.  
In our work, only one-body electromagnetic operators have been considered.

Experimentally, the g.s. binding energy of $^{18}$Ne is -132.142 MeV, and the INOY interaction underbinds the g.s. by 1.187 MeV. The other two realistic interactions underestimate the binding energy of $^{18}$Ne more significantly.  The binding energies of $^{18}$Ne obtained for all the three interactions are better than the HFB and the PGCM results. However, the IM-NCSM result is close to the experimental binding energy and overbinds $^{18}$Ne by 637 keV. Similarly, the calculated g.s. energy of $^{19}$Ne obtained from INOY interaction underbinds the g.s. by 1.196 MeV which is better than the CDB2K and N$^3$LO results.  The binding energy of $^{20}$Ne for INOY interaction is particularly good and underbinding it only by 100 keV compared to the experimental data. This result is even better than the IM-NCSM result reported in Table \ref{em} which overbinds $^{20}$Ne by 1.865 MeV. Fig. \ref{g.s.} shows the mass number dependence of the calculated g.s. energies for neon isotopes along with the experimental results. The NCSM results for g.s. energies are taken for the maximum possible $N_{max}$ at corresponding optimal frequencies. From the figure, it is seen that the g.s. energies of $^{18-20}$Ne isotopes obtained from INOY interaction are close to the experimental results, while the other two interactions significantly underbind the g.s of these nuclei. For $^{21-24}$Ne, the INOY interaction overbinds the corresponding g.s compared to the experimental data.  The CDB2K and N$^3$LO inteactions underbind the g.s. of $^{19-24}$Ne by about 11.8-14.10 MeV and 8.55-10.1 MeV, respectively. Overall, the INOY interaction is better for g.s. energy calculations of $^{18-20}$Ne isotopes due to its phenomenological non-local character at small distances. More binding energy due to non-local $NN$ potential is explained in Ref. \cite{CDB2K2} for the case of triton binding energy.  The IM-NCSM results for even neon {\color{black} isotopes} are better than all the three interactions we considered which underbind these isotopes. IM-NCSM method reproduces the g.s. energy well
except for $^{30}$Ne, which is observed to be less bound than $^{28}$Ne \cite{Ne5}.

\begin{figure}
	\begin{center}
	\includegraphics[width=9cm]{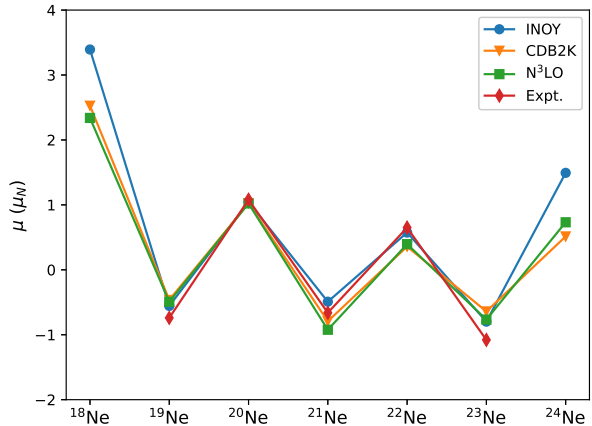}
		\end{center}
	\caption{Comparison between calculated magnetic moments at highest $N_{max}$ using  INOY, CDB2K, and N$^3$LO  $NN$ interactions with the experimental results. }\label{m.m.}
\end{figure}

Among the electromagnetic properties, the $E2$ transition strengths and moments of the Ne isotopes are not in good agreement with the experimental data as shown in the Table \ref{em}. 
We have noted that the rate of convergence of $E2$ observables is slow compared to the convergence of excitation energies. The $E2$ operators contain $r^2$ dependence term, making those observables sensitive to the long-range part of the nuclear wave functions. In order to get correct asymptotic behaviour of nuclear wave functions and converged results for $E2$ observables, large $N_{max}$ basis spaces are needed. The N$^3$LO interaction results are better for $E2$ observables than the other two interactions. Still, for N$^3$LO, we only get about half the experimental $E2$ transition strengths, and moments for neon isotopes {\color{black} except}  $^{20}$Ne. Similar kind of observations for $E2$ transition strengths was also reported in Ref. \cite{Roth-2016} for $^{12}$C and $^6$Li using the importance truncated no-core shell model (IT NCSM).
Unlike the case of $E2$ observables, the $M1$ observables do not depend on the spatial distance and show different convergence behaviour. The $M1$ observables are dependent only on the spin and angular momentum of nuclear states. 
Fig. \ref{m.m.} shows the magnetic moments of the studied neon isotopes. All three interactions well reproduce the general trend of experimental magnetic moments.

 \begin{table}
    \caption{\label{em} The g.s. energies and electromagnetic observables of $^{18-24}$Ne corresponding to the largest $N_{max}$ at their optimal HO frequencies. The g.s. energies, quadrupole moments, magnetic moments, $E2$ and $M1$ transitions are in MeV, barn (b), nuclear magneton ($\mu_{N}$), $e^{2}$ fm$^{4}$ and $\mu_{N}^{2}$, respectively. Experimental values are taken from Refs. \cite{NNDC,Qandmag}.  We have also compared our results with the  results available in the literature from the Refs.\cite{Ne5,Ne5**}}.
    \resizebox{\textwidth}{!}{%
    \begin{tabular}{cccccccccc}
    \hline 
    $^{18}$Ne  & EXPT  & INOY & CDB2K  & N$^3$LO & VS-IMSRG  &  HFB  & PGCM   & IM-NCSM  & BMBPT(3)    \tabularnewline
    \hline 
    Q($2^{+}$)    & NA & -0.014 & -0.087 & -0.012 & - & -&  -0.141 & -0.04 & - \\
    $\mu$($2^{+}$) & NA & 3.393 & 2.527 & 2.339   & - & -&  2.469 &  3.07 &- \\ 
		E$_{g.s.}$($0^{+}$)& -132.142 & -130.955 & -113.333 & -115.057  & -& -75.448 &  -79.228 & -132.78 & -126.824 \\ 
		$B(E2;2_{1}^{+}$ $\rightarrow$ $0_{1}^{+}$) & 49.6(50) & 7.097 & 19.727 & 25.143  & 19.0 & -&  55.00 & 17.93 & -\\ 
		$B(E2;4_{1}^{+}$ $\rightarrow$ $2_{1}^{+}$) & 24.9(34)  & 5.614 & 13.566 & 17.260  &-&-&-&-&-\\ 
		\hline 
		$^{19}$Ne & EXPT  & INOY & CDB2K  & N$^3$LO & VS-IMSRG  &  HFB  & PGCM   & IM-NCSM  & BMBPT(3)    \tabularnewline 
		\hline \vspace{-2.8mm}\\
		Q($5/2_1^{+}$) & NA & -0.059 & -0.078  & -0.089 &-& -&-& -&-\\ 
		$\mu$($5/2_1^{+}$) & -0.740(8) & -0.556 & -0.464 & -0.492  &-& -&-& -&-\\ 
		E$_{g.s.}$($1/2^{+}$)& -143.780  & -142.584 & -129.948 & -133.522  &-& -&-& -&-\\ 
		$B(E2;5/2_1^{+}$ $\rightarrow$ $1/2_{1}^{+}$) & 39.7(15) & 8.868 & 15.543 & 20.044 & 25.0&-& -&-& - \\ 
		$B(M1;3/2_{1}^{+}$ $\rightarrow$ $5/2_{1}^{+}$) & 0.89(50)  & 2.826 & 2.867  & 2.848 &-& -&-& -&- \\ 
		\hline 
		$^{20}$Ne & EXPT  & INOY & CDB2K  & N$^3$LO & VS-IMSRG  &  HFB  & PGCM   & IM-NCSM  & BMBPT(3)    \tabularnewline 
		\hline \vspace{-2.8mm}\\
		Q($2^{+}$) & -0.23 & -0.065  & -0.082  & $<$ 0.001 &- & -& -0.183 & -0.011 & - \\ 
		$\mu$($2^{+}$) & 1.08 & 1.044  & 1.025 & 1.022     &- &- & 1.003  &1.04 &-\\ 
		E$_{g.s.}$($0^{+}$)& -160.645 & -160.545 & -146.558 & -149.939  & -& -94.259 &-101.675 & -162.510 & -152.648\\ 
		$B(E2;2_{1}^{+}$ $\rightarrow$ $0_{1}^{+}$) & 65.4(32) & 10.346 & 16.182 & $<$ 0.001 & - & -& 80.895 & 27.55 &-\\ 
		$B(E2;4_{1}^{+}$ $\rightarrow$ $2_{1}^{+}$) & 71.0(64)  & 12.567  & 19.964 & $<$ 0.001  &-& -&-& -&-\\ 
		\hline 
		$^{21}$Ne & EXPT  & INOY & CDB2K  & N$^3$LO & VS-IMSRG  &  HFB  & PGCM   & IM-NCSM  & BMBPT(3)    \tabularnewline 
		\hline \vspace{-2.8mm}\\
		Q($3/2_1^{+}$) & 0.1016(8) & 0.045 & 0.055 & 0.060  & - & - & -& - & -\\ 
		$\mu$($3/2_1^{+}$) & -0.66170(3) & -0.491 & -0.792 & -0.923  & - & -& -& -&-\\ 
		E$_{g.s.}$($3/2^{+}$)& -167.406  & -170.655 & -153.819 & -156.790 & - &- & -& -&-\\ 
		$B(E2;9/2_{1}^{+}$ $\rightarrow$ $5/2_{1}^{+}$) & 54.0(75) & 8.545 & 14.336 & 19.720 & - & -& -&- &-\\ 
		$B(M1;5/2_{1}^{+}$ $\rightarrow$ $3/2_{1}^{+}$) & 0.1275(25) & 0.202 & 0.151 & 0.167 & - & -& -&- &-\\ 
		\hline 
		$^{22}$Ne & EXPT  & INOY & CDB2K  & N$^3$LO & VS-IMSRG  &  HFB  & PGCM   & IM-NCSM  & BMBPT(3)    \tabularnewline 
		\hline \vspace{-2.8mm}\\
		Q($2^{+}$) & -0.215 & -0.060  & -0.075  & -0.086  & - & - & -0.168 &  -0.096 & -  \\ 
		$\mu$($2^{+}$) & 0.65(2)  & 0.574 & 0.358 & 0.395 & - & - & 0.774  &  0.56 &  -\\ 
		E$_{g.s.}$($0^{+}$)& -177.770 & -182.675 & -163.860 & -166.865  &  & -103.158& -109.027& -179.580& -168.122\\ 
		$B(E2;2_{1}^{+}$ $\rightarrow$ $0_{1}^{+}$) & 46.72(66) & 8.813 & 13.650 & 17.668  &  22.7 & - & 70.007 & 22.52 & - \\ 
		$B(E2;4_{1}^{+}$ $\rightarrow$ $2_{1}^{+}$) & 64.1(14) & 11.856  & 18.001 & 21.011 & - & - &- &- & -\\ 
		\hline
      $^{23}$Ne  & EXPT  & INOY & CDB2K  & N$^3$LO & VS-IMSRG  &  HFB  & PGCM   & IM-NCSM  & BMBPT(3)    \tabularnewline 
		\hline 
		Q($5/2^{+}$) & 0.145(13) & 0.070 & -0.045 & -0.066 &  -&  -& -& -&  -\\ 
		$\mu$($5/2^{+}$) & -1.0794(10) & -0.798 & -0.641 & -0.766  & - & - & -& -& -\\ 
		E$_{g.s.}$($5/2^{+}$)& -182.971 & -190.836 & -169.836 & -173.163 & - & - &- & -& -\\ 
		$B(M1;3/2_{1}^{+}$ $\rightarrow$ $5/2_{1}^{+}$) & NA & 0.007 & 0.034  & 0.098  & - & - & -& -& -\\ 
		\hline 
		$^{24}$Ne  & EXPT  & INOY & CDB2K  & N$^3$LO & VS-IMSRG  &  HFB  & PGCM   & IM-NCSM  & BMBPT(3)    \tabularnewline 
		\hline \vspace{-2.8mm}\\
		Q($2^{+}$)     & NA & -0.011 & -0.030 & -0.068  & - & -&-0.160  &-0.023  & -    \\ 
		$\mu$($2^{+}$) & NA & 1.493  & 0.514  & 0.732   & - & -& 1.011   & 0.99    & -  \\ 
		E$_{g.s.}$($0^{+}$)& -191.840 & -202.234 & -179.995 & -183.290 & -  & -107.923 & -113.283 & -193.34& -177.572\\ 
		$B(E2;2_{1}^{+}$ $\rightarrow$ $0_{1}^{+}$) & 28.0(66) & 6.204 & 10.282 & 16.155  &  13.8 & - & 59.91& 16.69 & -\\ 
		$B(E2;4_{1}^{+}$ $\rightarrow$ $2_{1}^{+}$) & NA & 7.254 & 9.883 & 13.406 & -& - & -& -& -\\ 
		\hline
		\hline
    \end{tabular} }
\end{table}

 In the present work we have also compared our no-core shell model results with other \textit{ab initio} results available in the literature Refs. \cite{Ne5,Ne5**}. 
	The $B(E2;2_{1}^{+}$ $\rightarrow$ $0_{1}^{+}$) for $^{18}$Ne with VS-IMSRG, IM-NCSM and PGCM are 19.0, 17.93 and 55.00 $e^2fm^4$, respectively. While the VS-IMSRG result is close to the CDB2K result, the N$^3$LO result is slightly better than the IM-NCSM result. The  E2 strength with PGCM for the $^{18}$Ne is  large by a factor of 1.11 compared to the experimental data.
	In the case of $^{19}$Ne, the $B(E2;5/2_1^{+}$ $\rightarrow$ $1/2_{1}^{+}$) is 25 $e^2fm^4$ with VS-IMSRG, better than INOY, CDB2K and N$^3$LO results.
	In the case of $^{20}$Ne, the calculated $B(E2;2_{1}^{+}$ $\rightarrow$ $0_{1}^{+}$) with PGCM is large in comparison to the experimental data, while with IM-NCSM it is small.
	The $B(E2;2_{1}^{+}$ $\rightarrow$ $0_{1}^{+}$) for $^{22}$Ne ($^{24}$Ne) with N$^3$LO, IM-NCSM and VS-IMSRG are too small by a factor of 0.38 (0.58), 0.49 (0.49) and 0.48 (0.60), respectively compared to the experimental data. While the E2 strengths of  $^{22,24}$Ne obtained for PGCM is large compared to the experimental data, they are significantly small for INOY and CDB2K interactions. So, our NCSM results of E2 strengths with N$^3$LO are consistent with other ab {\color{black} initio} results. Similarly, the quadrupole moments of $^{20-23}$Ne {\color{black} isotopes} obtained from our NCSM study along with the results of IM-NCSM are small compared to the available experimental data.

\begin{figure}
	 \includegraphics[width=8.8cm,height=5.6cm]{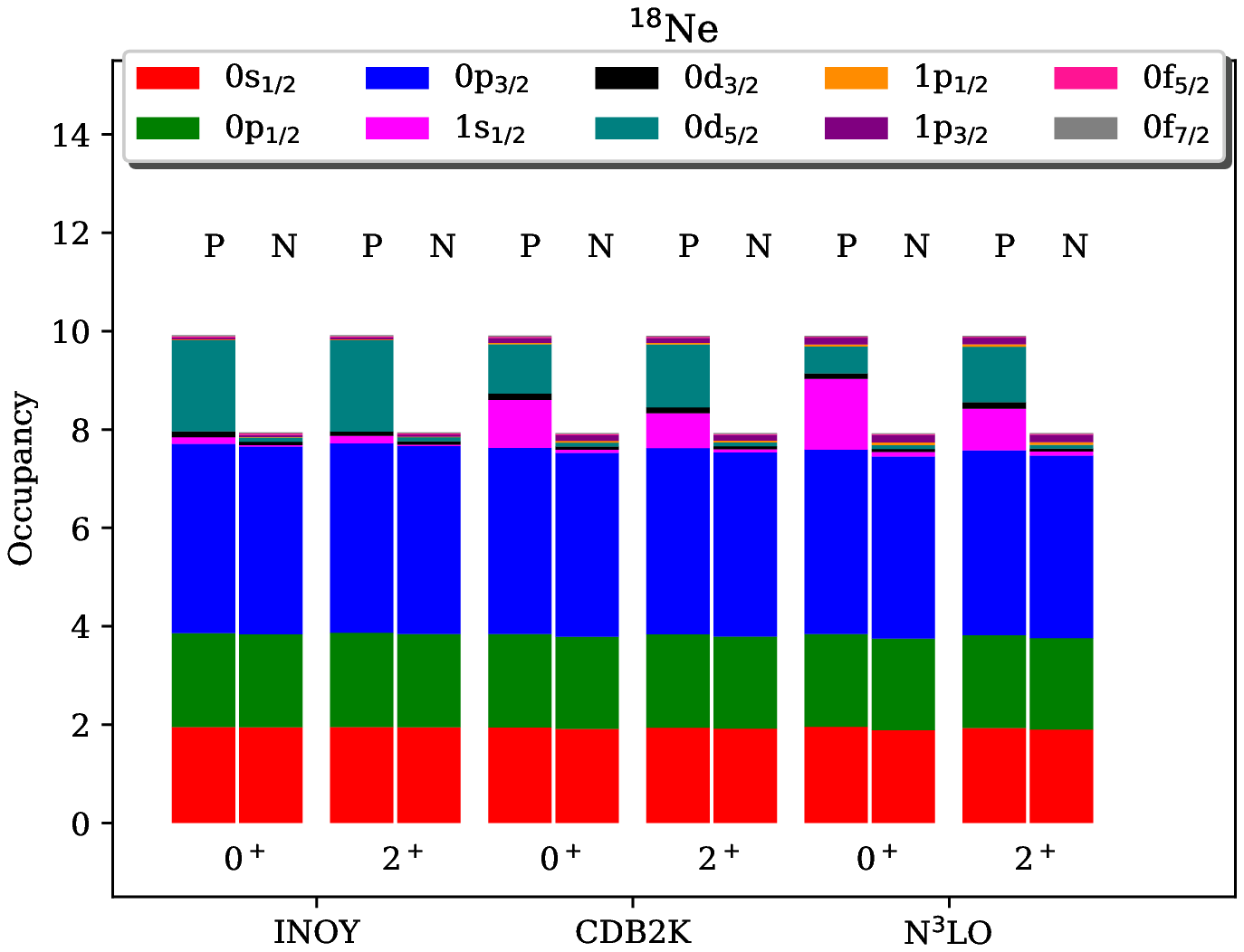}
	 \includegraphics[width=8.8cm,height=5.6cm]{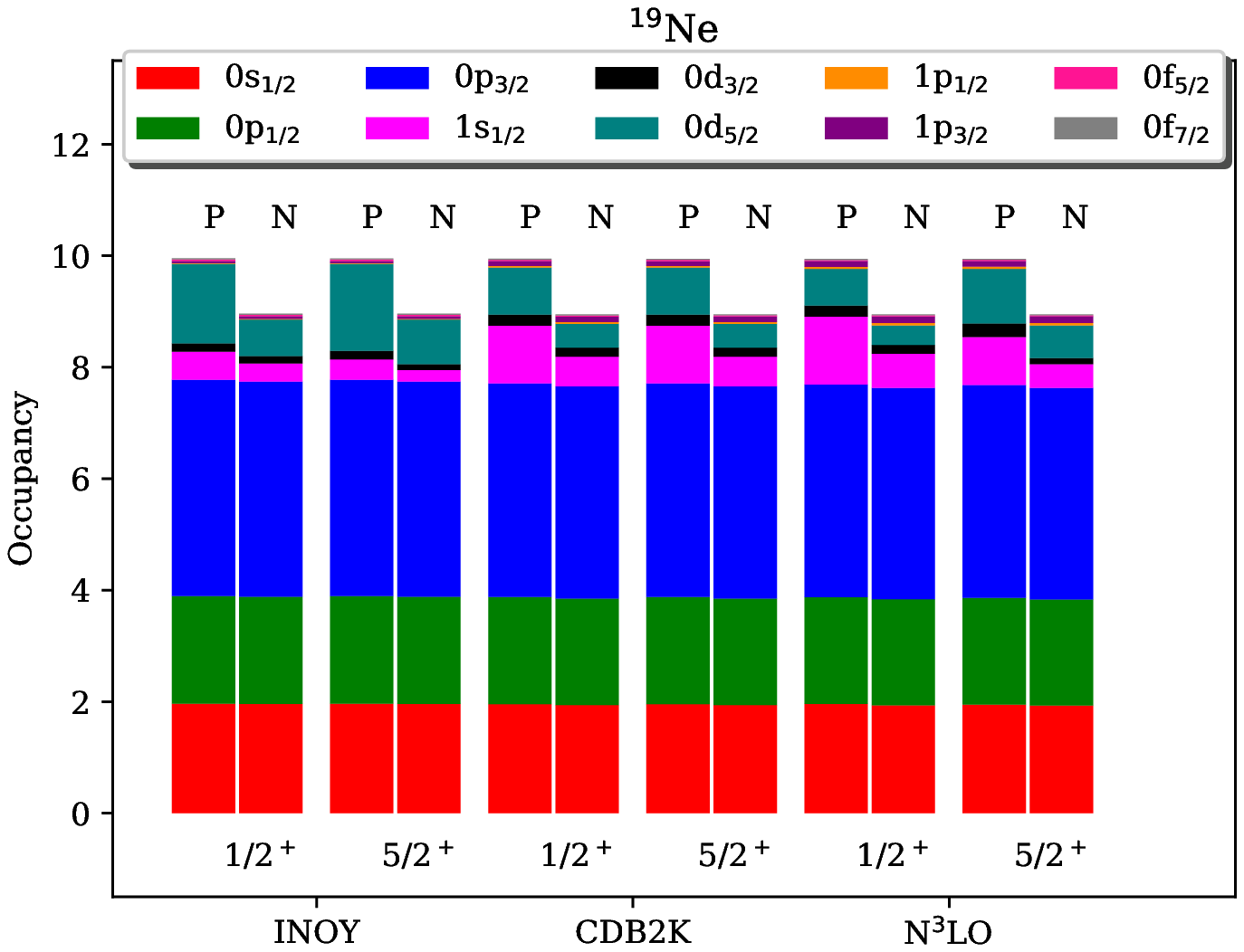}
	 \includegraphics[width=8.8cm,height=5.6cm]{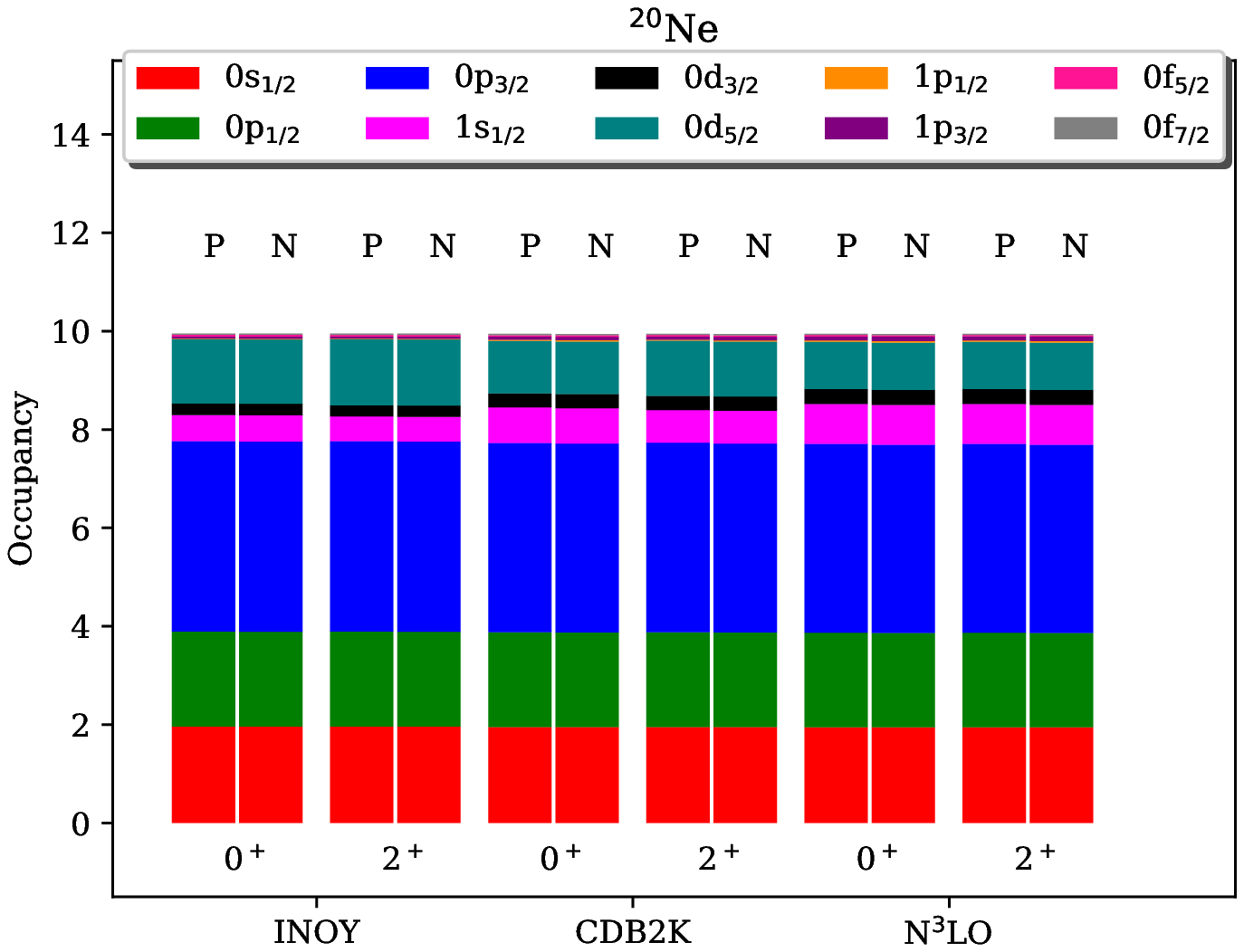}
	 \includegraphics[width=8.8cm,height=5.6cm]{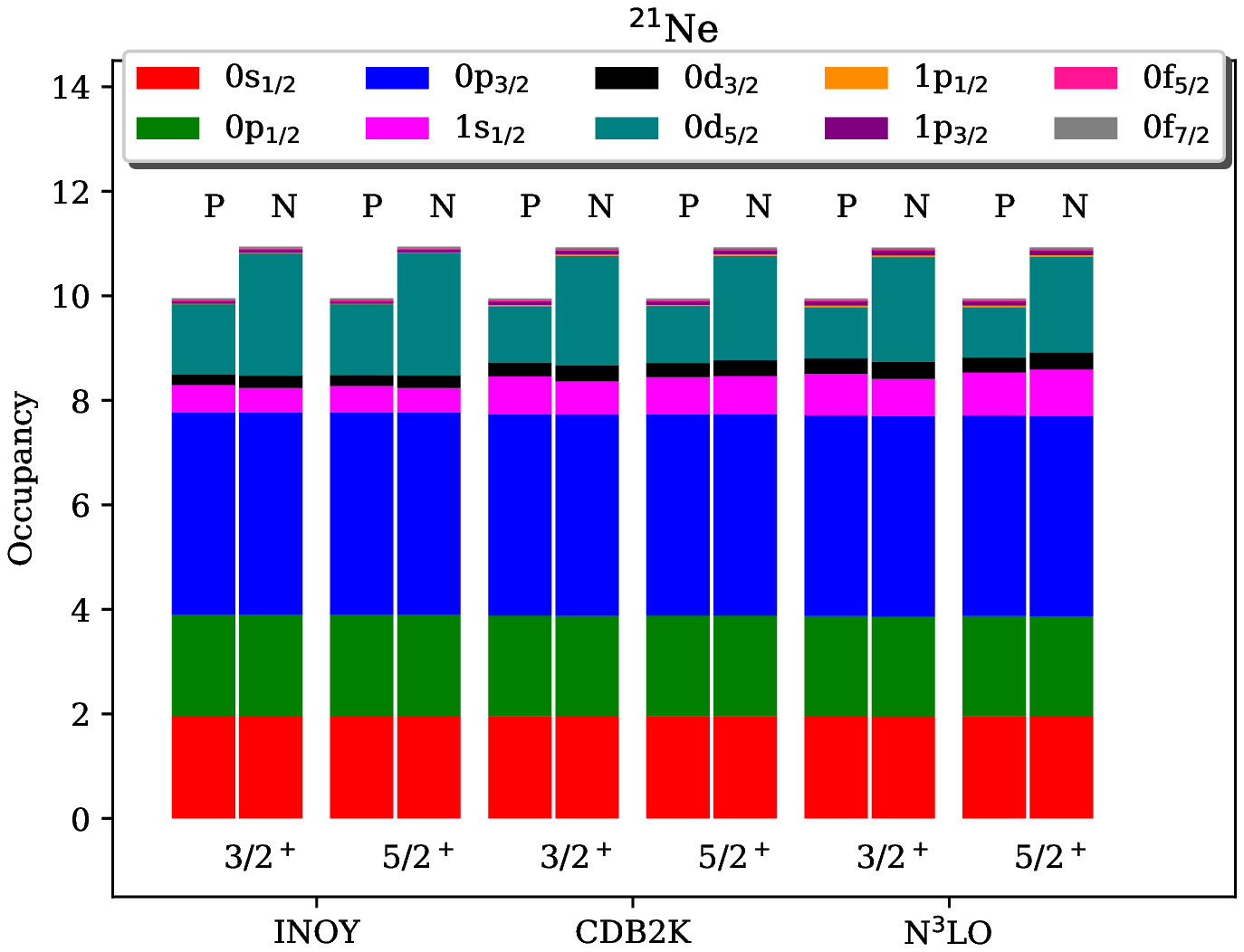}
    \includegraphics[width=8.8cm,height=5.6cm]{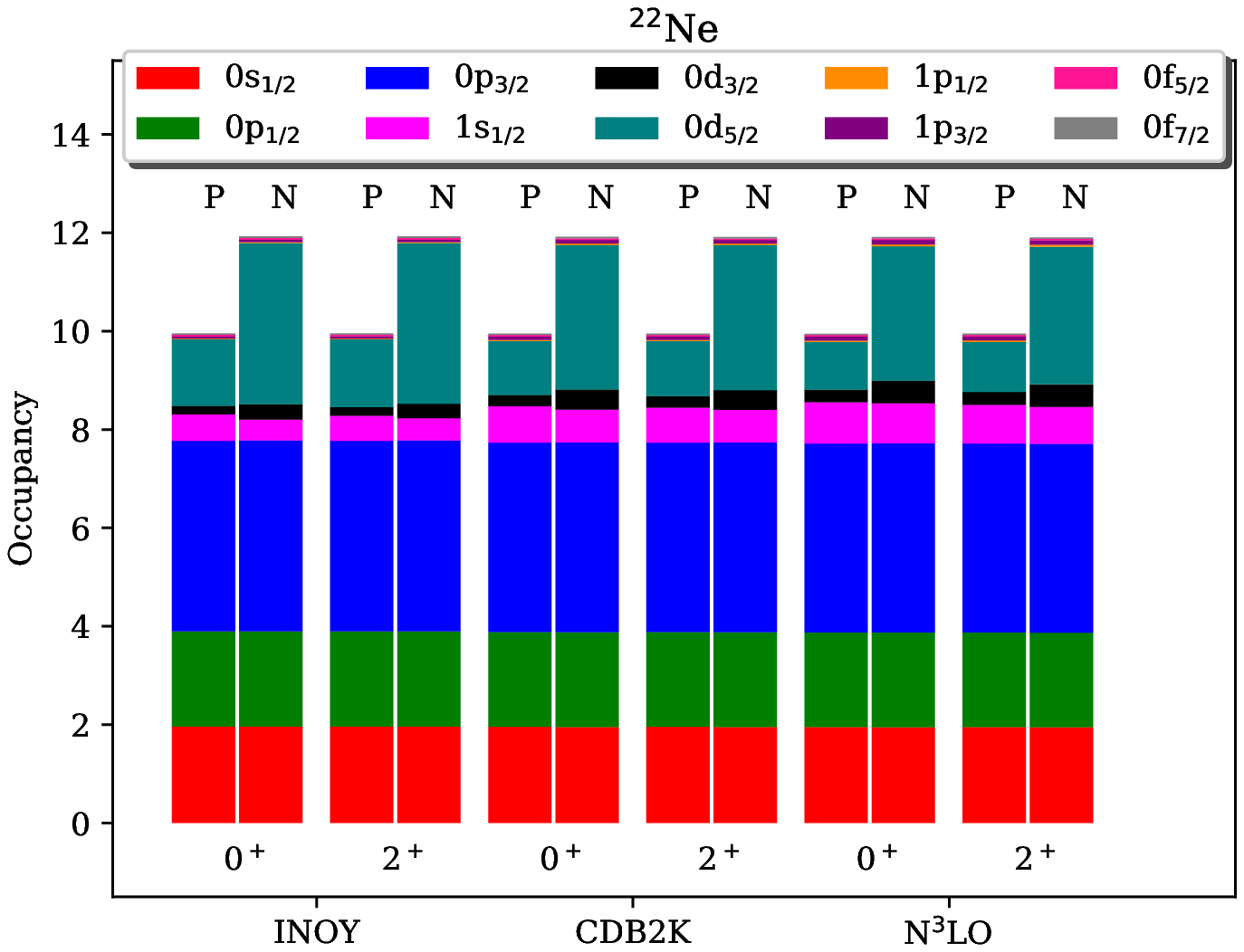}
    \includegraphics[width=8.8cm,height=5.8cm]{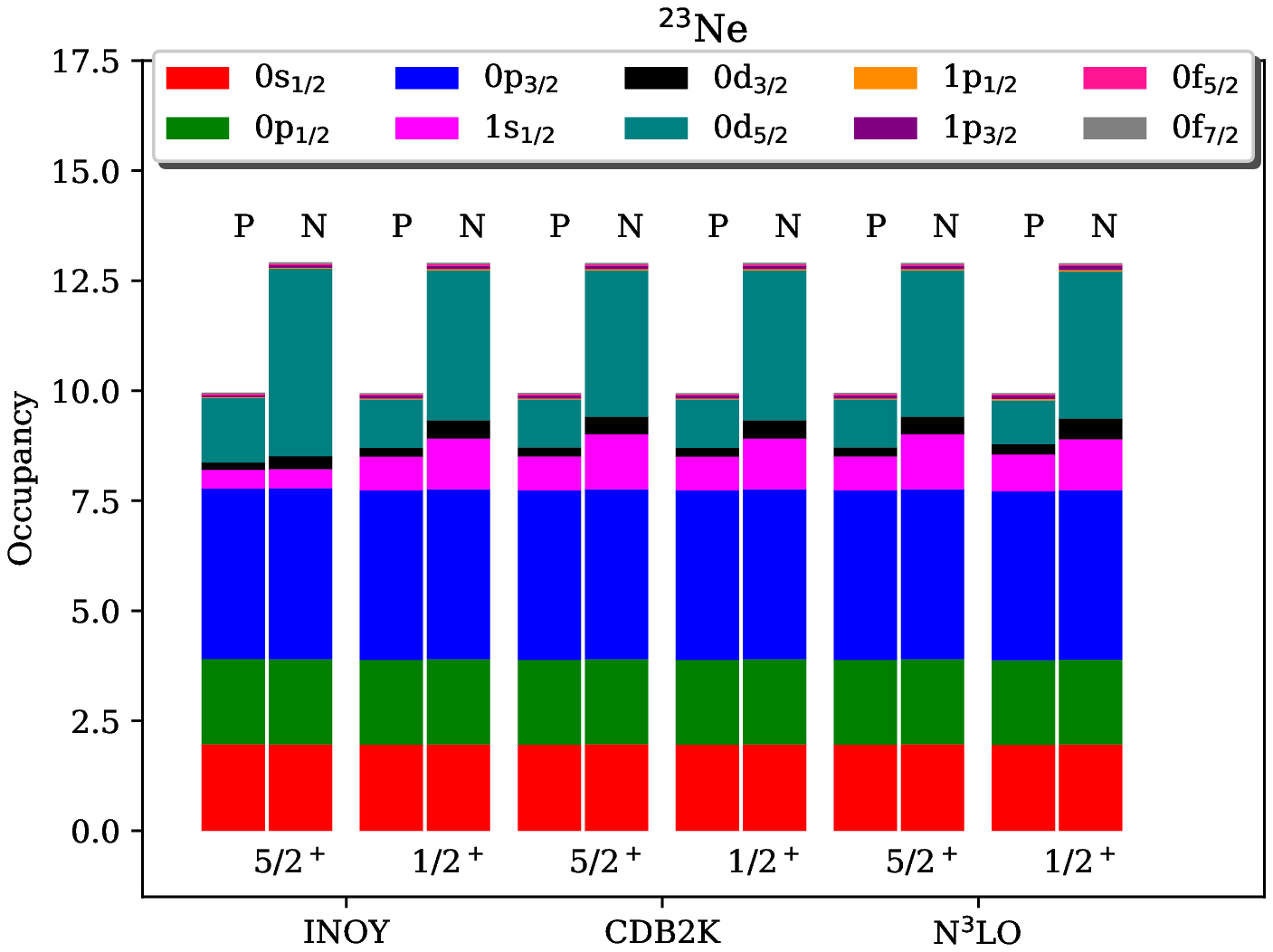}
   \hspace{10cm}  ~~~~~~~~~
   \vspace{-1cm}
 	 	  \begin{center}
 	 	  \includegraphics[width=8.8cm,height=5.6cm]{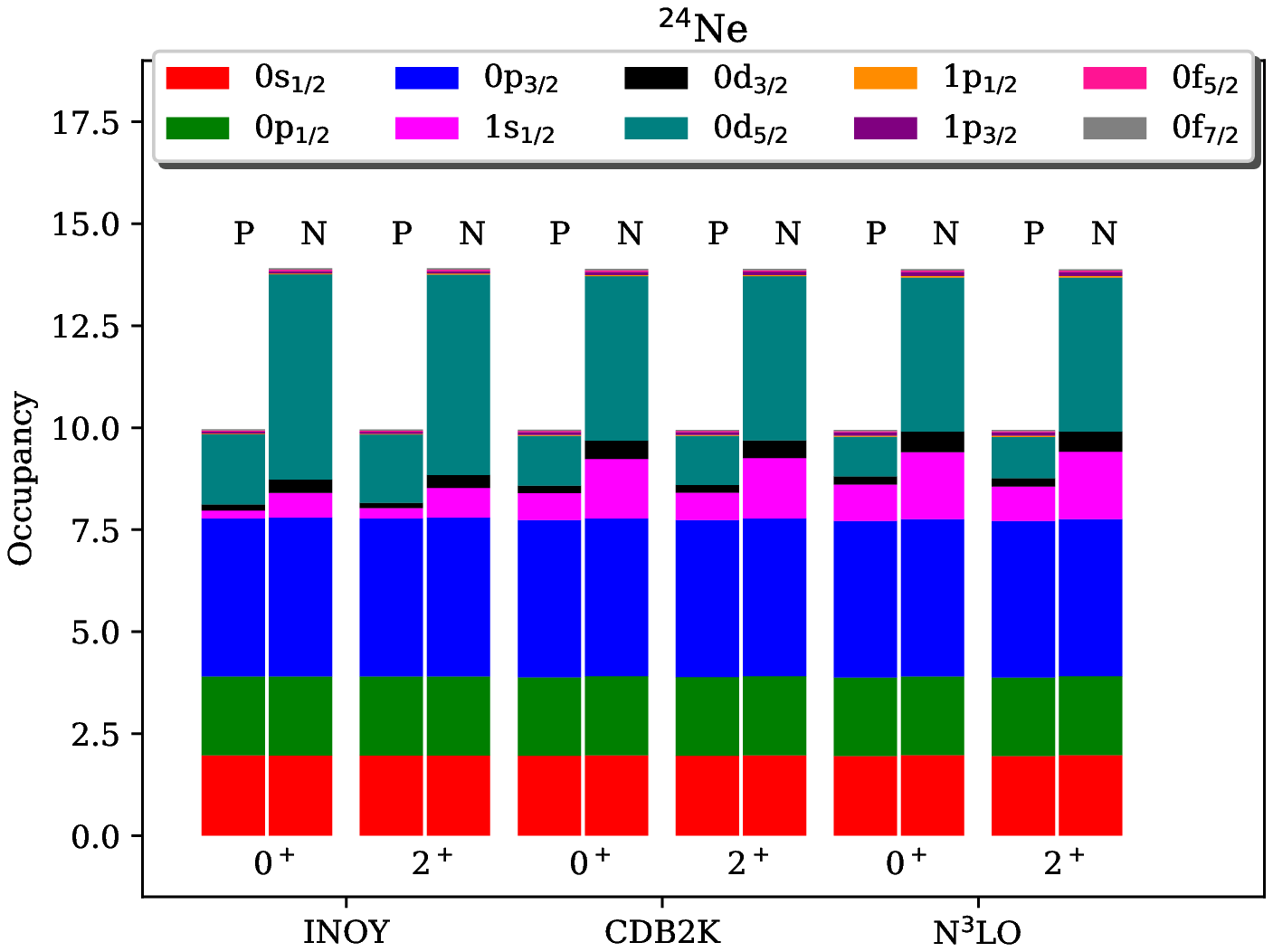}	   
	\hspace{-4cm}\caption{ Occupation number for ground and first excited states of $^{18-24}$Ne isotopes with the INOY, CDB2K and N$^3$LO interactions. ‘P’ and ‘N’ represent proton and neutron occupation numbers, respectively.
 }
 \label{occupancy}
 \end{center}
 
\end{figure}

  To know details about the role of the different orbitals in the NCSM wave-function we have plotted occupancy of g.s. and the first excited states of $^{18-24}$Ne isotopes in Fig. \ref{occupancy}.
Here we have compared the occupation number of different realistic interactions. We have reported only proton and neutron occupation numbers up to $fp$-shell, beyond this it is very small. For $^{18}$Ne, the occupancy of both proton and neutron $p_{3/2}$ orbital is dominant for $0^+$ and $2^+$ states.
As we move from $^{18}$Ne to $^{24}$Ne the occupancy of $\nu d_{5/2}$ orbital is increasing.
 Also, we noticed a significant difference in  occupancies of $\pi 1s_{1/2}$ and $\nu 1s_{1/2}$ for INOY and other two {\color{black} interactions} for a particular neon isotope. The differences in the {\color{black} occupancies} of these orbitals for all the three interactions are reflected in the results of the electromagnetic properties of different neon isotopes using these interactions. 

In the present work, we considered only one-body electromagnetic operators but, for $BE2$ like non-scalar operators the induced two-body terms dominate over the corresponding one-body term. That is one source of the missing $BE$2 strengths and moments in our NCSM results. Also, for some of the neon isotopes correlation effects due to deformation should be included. These two factors are not considered in our study of neon isotopes.

\section{Point-proton radii}
\label{sect6}
In Table \ref{rp}, we have presented the results of point-proton radii ($r_p$) obtained from INOY, CDB2K, and N$^3$LO interactions along with the experimental values for $^{18-24}$Ne isotopes. 
These NCSM results for $r_p$ are reported corresponding to highest possible $N_{max}$ basis at their 
respective optimal frequencies.
The root mean square point-proton distribution ($r_p$) is a long-range operator like the $E2$ transition operator, which is sensitive to the long-range part of nuclear {\color{black} wave functions}. 
The experimental values of charge-radii are taken from the ADNDT2013 compilation Ref. \cite{charge_radii}. To calculate the point-proton radius from the experimental charge radius, the following formula is used:
\begin{equation}
\langle r^2 \rangle_{p} = \langle r^2 \rangle_{c} - \langle R_p^2 \rangle - \frac{N}{Z} \langle R_n^2 \rangle - \frac{3}{4 m_p^2}\\
\end{equation}

where, $\langle R_p^2 \rangle$ and $\langle R_n^2 \rangle$ are the squared charge radius of proton  and neutron, respectively and the last term is the Darwin-Foldy term related to the relativistic correction in natural units. These values are taken to be $\langle R_p^2 \rangle^{1/2}$ = 0.8783(86) fm, $\langle R_n^2 \rangle$ = -0.1149(27) fm$^2$ \cite{charge_radii} and 3/(4$m_p^2$) = 0.033 fm$^2$ \cite{C.Forseen}. The squared point-proton radius relative to the center of mass of all nucleons is evaluated with 
\begin{equation}
  r_p^2  = \frac{1}{Z} \sum_{i=1}^{Z}  \vec{r_i} - \vec{R}_{CM}^2.
\end{equation}
The operator in the above equation is a two-body operator. It is reduced to a more suitable form involving one-body and two-body operators to evaluate two-body matrix elements for this operator. After finding the two-body matrix elements, the expectation value of the $r_p$ operator is similar to the calculations of the ground-state energies.  

\begin{table}[ht]
	\centering
	\caption{\label{rp} The point-proton radii ($r_p$) calculated with INOY, CDB2K and N$^3$LO. These are given in fm.}
		\begin{tabular}{ccccc}
			\hline
			\hline 
			$r_p$ & Expt. & INOY & CDB2K & N$^3$LO\\
			\hline \vspace*{.5mm}
			$^{18}$Ne & 2.8490(84) & 2.21 & 2.53 & 2.69\\
			$^{19}$Ne & 2.8893(49) & 2.19 & 2.40 & 2.53\\
			$^{20}$Ne & 2.8885(34) & 2.18 & 2.39 & 2.53\\
			$^{21}$Ne & 2.8530(43) & 2.18 & 2.39 & 2.53\\
			$^{22}$Ne & 2.8374(49) & 2.17 & 2.39 & 2.53\\
			$^{23}$Ne & 2.7956(79) & 2.16 & 2.38 & 2.52\\
			$^{24}$Ne & 2.7876(86) & 2.16 & 2.38 & 2.53\\
			\hline
			\hline	
		\end{tabular}
\end{table}

\begin{figure}
	\includegraphics[width=8.8cm]{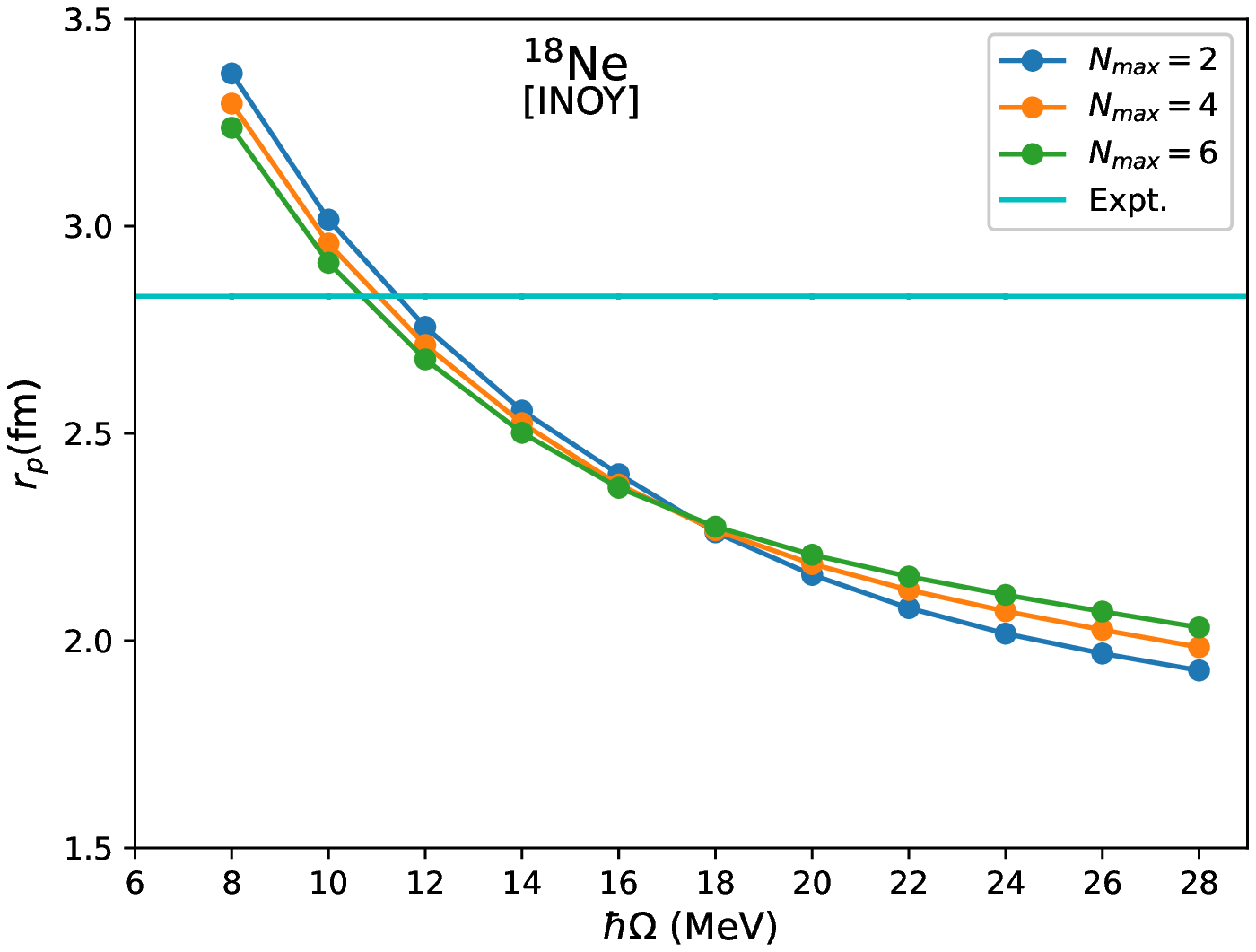}
	\includegraphics[width=8.8cm]{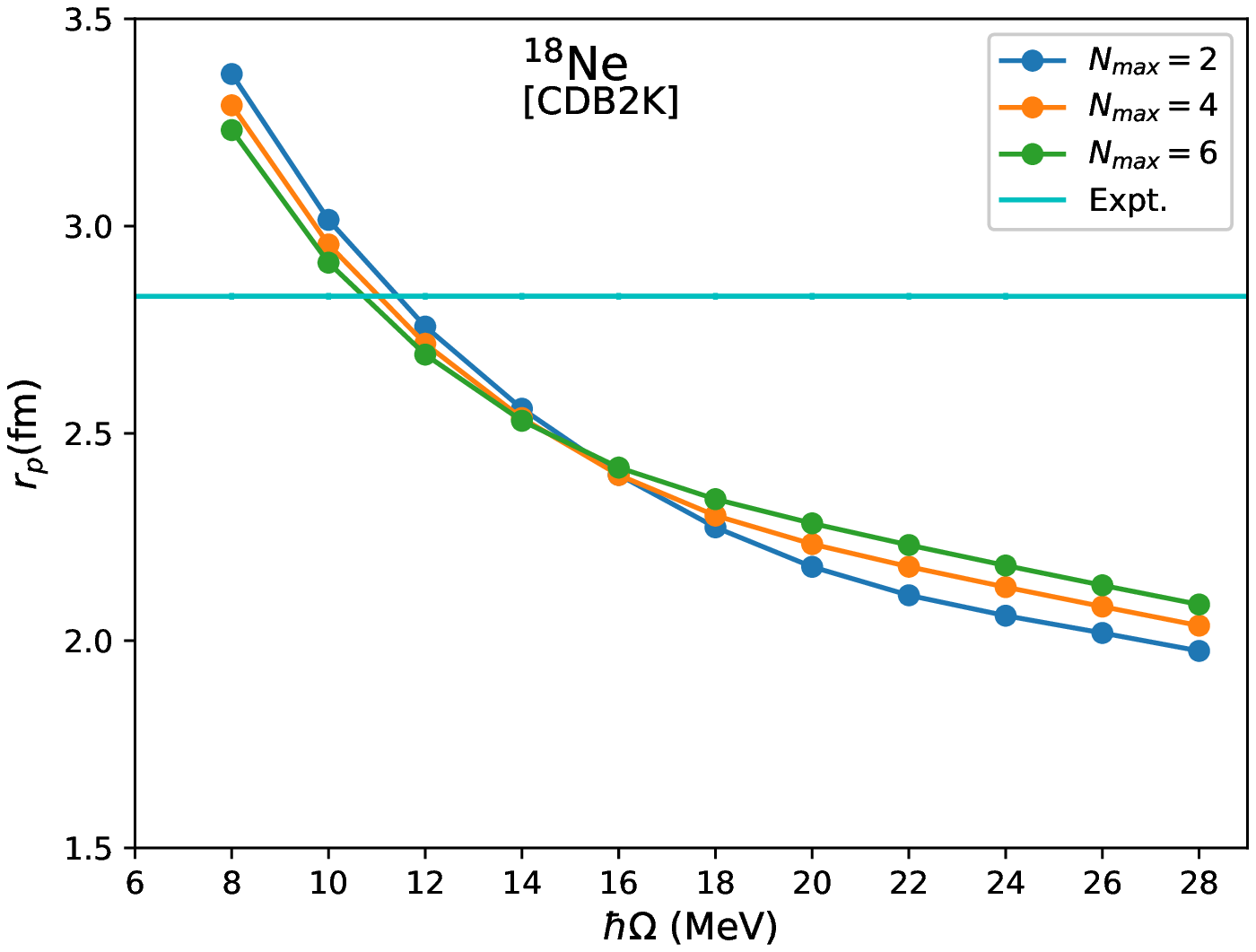}
	\begin{center}
	\includegraphics[width=8.8cm]{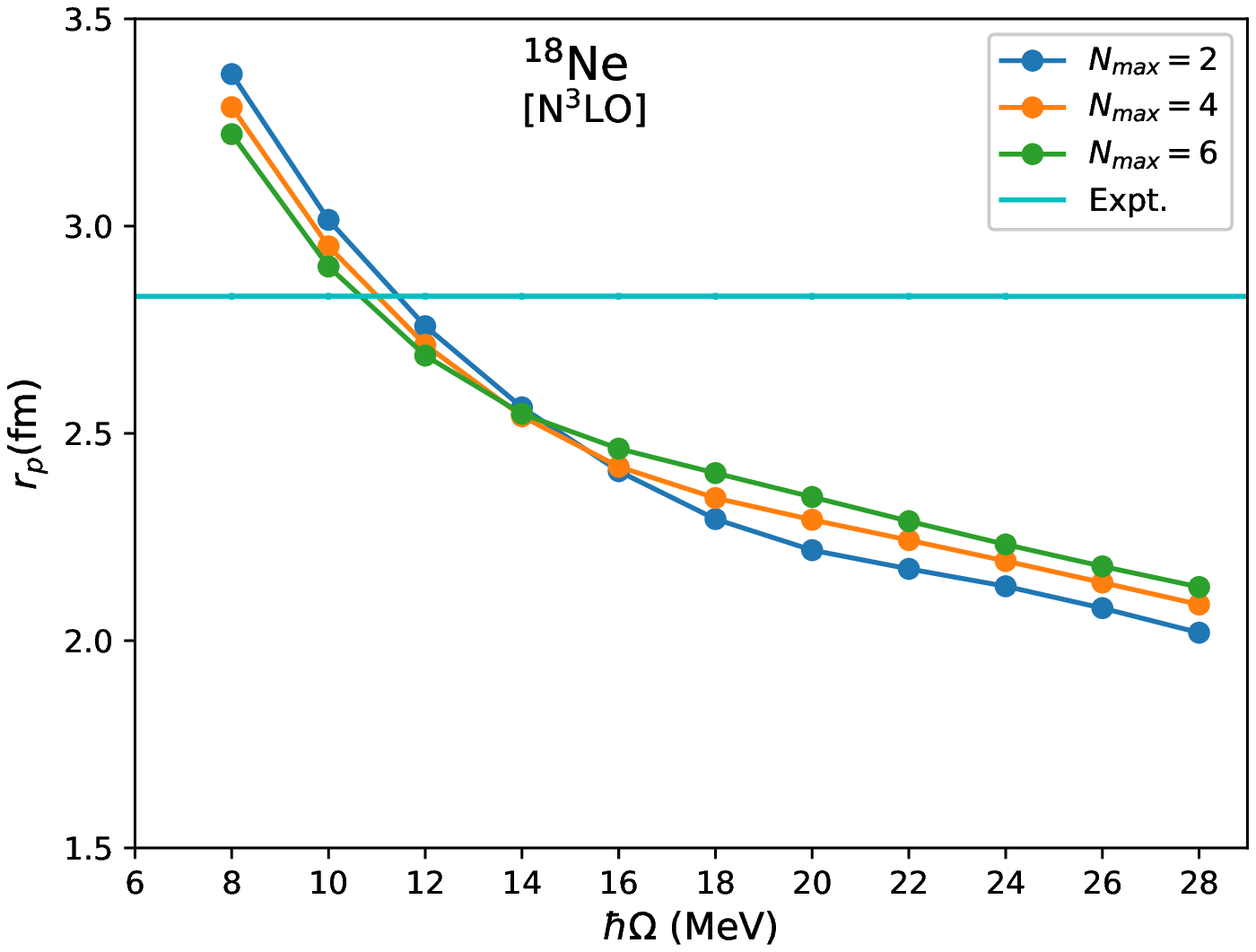}
	\end{center}
	\caption{Point-proton rms radius of $^{18}$Ne g.s as a function of $\hbar \Omega$. }\label{point proton radius1}
\end{figure}

\begin{figure}
\begin{center}
	\includegraphics[width=9cm]{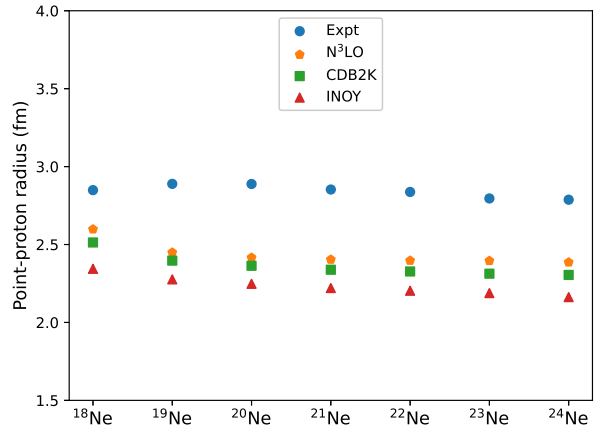}
	\end{center}
	\caption{Comparison of the NCSM results of point-proton rms radius with the INOY, CDB2K and N$^3$LO $NN$ interactions to the experimental data.} \label{rp_scatter}
\end{figure}

The $r_p$ dependence on $N_{max}$ and $\hbar \Omega$ for $^{18}$Ne isotope corresponding to INOY, CDB2K and N$^3$LO interactions are shown in Fig. \ref{point proton radius1}. From the figure, it can be seen that the curves of $r_p$ as a function of $\hbar \Omega$ for different $N_{max}$ appear to
intersect at a common point for all the three interactions. For large $N_{max}$ calculations, the $\hbar \Omega$ corresponding to this crossing point should be smaller than the variational minimum of the g.s. energy \cite{charge_radii4}. It is observed that at a lower HO frequency, the calculated $r_p$ decrease with increasing $N_{max}$, while at a high HO frequency, it increases with increasing $N_{max}$. A region of calculated point-proton radii around the crossing point is independent of $N_{max}$. That crossover can be taken as a true converged point-proton radius \cite{charge_radii4, charge_radii3}. 
In this work, the intersection point of two $r_p$ curves corresponding to two successive highest $N_{max}$ are taken as the converged point-proton radius. In the first panel of Fig. \ref{point proton radius1}, the $r_p$ curves corresponding to $N_{max}$= 4 and $N_{max}$= 6 for INOY interaction cross each other approximately at 2.34 fm and this value is considered to be the point-proton radius obtained from INOY interaction. Similarly, the point-proton radii of $^{18}$Ne obtained from CDB2K, and N$^3$LO are 2.51 fm and 2.60 fm, respectively. The point-proton radii for other neon isotopes are found similarly by calculating  the intersection points of $N_{max}$ = 2 and $N_{max}$ = 4 curves corresponding to different $NN$ interactions. The converged point-proton radii of $^{18-24}$Ne isotopes obtained from INOY, CDB2K and N$^3$LO interactions are shown in Fig. \ref{rp_scatter} along with the experimental $r_p$ extracted from Ref. \cite{charge_radii}.  
From this figure, it can be seen that all the three interactions show the general trend of monotonic decrease of experimental point-proton radii from $^{19-24}$Ne. However, the calculated converged $r_p$ is less than the experimental data. Among the three $NN$ interactions, the CDB2K and N$^3$LO give better $r_p$ values than those obtained from the INOY interaction. This observation can be correlated to the calculated g.s. energies shown in Fig. \ref{g.s.}, that is, more binding of g.s. correspond to a smaller radius.

 For the case of $^{18}$Ne, optimal $\hbar \Omega$ obtained for g.s. energy using INOY interaction is 20 MeV, whereas it is obtained to be $\hbar \Omega  \approx$ 16 MeV for g.s point-proton radius. For CDB2K and N$^3$LO interactions, the optimal $\hbar \Omega$ of  $^{18}$Ne g.s. energy and g.s. point-proton radii are almost same up to $N_{max}$ = 6 calculations. The converged $r_p$ of $^{18}$Ne obtained from N$^3$LO (up to N$_{max} = 6$) is almost 9\% less than the experimental $r_p$ value. For other neon isotopes, this difference in $r_p$ between the experimental and the calculated results from the same interaction is around 15\%, corresponding to NCSM calculations up to $N_{max}$ = 4 basis space. So, we expect that for higher $N_{max}$ calculations, the converged $r_p$ will be close to the experimental data. Also, the $\hbar \Omega$ corresponding to the converged $r_p$ will be small compared to the minimum of the variational ground-state energy.  
 
 From Fig. \ref{rp_scatter}, a dip in the experimental as well as in the calculated $r_p$ can be seen at $N$ = 14  suggesting a shell closure at that point. Also, the decrease in $r_p$ from $^{19}$Ne to $^{24}$Ne is an indication of a transition from deformation towards sphericity at the $N$ = 14 shell closure \cite{N=14}.

 Corresponding to other \textit{ab initio} results as reported in Ref. \cite{Ne5}, the rms charge radii are
obtained from IM-NCSM smaller than the experimental radii. The experimental radius is known up to
$^{28}$Ne. The trend obtained form the IM-NCSM calculations is similar to the experimental trend.

\section{Conclusions}
\label{sect7}
In this work, we have investigated spectroscopic properties of neon isotopes within \textit{ab initio} no-core shell model using realistic $NN$ interactions \textit{i.~e.} INOY, CDB2K and N$^3$LO. We have studied low-energy spectra, electromagnetic properties, and point-proton radii with aforementioned interactions. We observed that the binding energy of $^{18-20}$Ne obtained for INOY interaction is very close to the experimental g.s. binding energies, and for $^{21-24}$Ne isotopes, it overbinds the corresponding g.s.  The g.s. binding energies of $^{20}$Ne for INOY interaction is in excellent agreement with the experimental data with the difference of only 100 keV. Comparing with other \textit{ab initio} results, the g.s. energies of $^{22, 24}$Ne isotopes from IM-NCSM \cite{Ne5} are better than all the three interactions considered here.  The low-energy spectra obtained for all the three $NN$ interactions provide good agreement with the experimental data  except for $^{23}$Ne isotope.  For the case of $^{23}$Ne, the g.s. is  well reproduced only by the INOY interaction. Also, several low-lying states show significant improvement on increasing basis size N$_{max}$.  

Among the electromagnetic properties, the fast convergence of $M1$ observables is observed due to their independence on spatial coordinates. The $B(M1; 5/2^+_1 \to 3/2^+_1)$ of $^{21}$Ne obtained from all the three interactions are close to the experimental $M1$ transition strength. 
Also, all these interactions reproduce the correct trend of experimental magnetic moments. On the other hand, the convergence of $E2$ observables is challenging to achieve in small model space calculations due to their $r^2$-dependence. These observables are sensitive to the long-range {\color{black} behaviour} of nuclear {\color{black} wave functions}, and a large $N_{max}$ basis space is needed to obtain converged results. The N$^3$LO provides better results for $E2$ transition strengths among the three realistic interactions except for $^{20}$Ne isotope. However, those values are only half of the experimental transition strengths.  Similar kind of underestimation of $E2$ strengths and moments are also reported in Ref. \cite{Ne5} for IM-NCSM, while the PGCM approach overestimate those observables.

The point-proton radius ($r_p$) is also sensitive to the long-range part of nuclear {\color{black}wave functions} like the $E2$ observables, and to obtain the converged results for $r_p$, we used the ``crossing point'' method. We find out that the point-proton radii calculated using N$^3$LO and CDB2K interactions for the g.s. of neon isotopes are better than INOY interaction, which is attributed to the fact that both N$^3$LO and CDB2K underbind the g.s. For N$^3$LO interaction, almost 6 \% higher converged $r_p$ is obtained for $N_{max}$ = 6 compared to $N_{max}$ = 4 calculations. So, better results for $r_p$ can be expected for higher $N_{max}$ calculations. We have noticed that the converged $r_p$ decrease slowly from $^{19}$Ne to $^{24}$Ne making a dip at $^{24}$Ne. It reflects a transition from deformation towards spherical shape at the $N$ = 14 shell closure. This observation can also be seen from the decrease in the $B(E2;2^+_1 \to 0^+_1)$ for even-Ne isotopes from $^{20}$Ne to $^{24}$Ne in Table \ref{em}.  We also observe the difference in optimal frequencies for the g.s. energy and the point-proton radius of these neon isotopes.

As diagonalization of large-matrix becomes a significant computational challenge for {\color{black} \textit{ab initio}} NCSM, recent modifications of conventional NCSM like IT-NCSM \cite{Roth-2016,Roth-2009} and SA-NCSM \cite{Ne7,Ne5*,SANCSM_2206} may help to study the full neon chain up to neutron drip line nuclei in the future. Also, the challenges of obtaining converged results for long-ranged observables can be tackled by the indirect approaches mentioned in Refs. \cite{ Roth-2016,M.Caprio2019, M.Caprio2022,M.Caprio2022b}. 

\vspace{0.8cm}
\section*{Acknowledgments}

 We acknowledge financial support from SERB (India), CRG/2019/000556. 
 We would like to thank Prof. Petr Navr\'atil for providing us his $NN$ effective interaction code
and for his valuable comments.  We would also like to thank Christian Forss\'en for making {\color{black} available} the pANTOINE code. Also we thank M. Frosini for useful discussions.



\section*{References}
\begin{thebibliography}{99}
	
	
	
	
	
	
	\bibitem{NCSM_r2} B. R. Barrett, P. Navr\'atil, and J. P. Vary, \textit{Ab initio} no core shell model, 
	{Prog. Part. Nucl. Phys. {\bf 69}, 131 (2013).}
	
	
    \bibitem{Maris} P. Maris, E. Epelbaum, R. J. Furnstahl, J. Golak, K. Hebeler, T. H\"uther, H. Kamada, H. Krebs, Ulf-G. Mei\ss{}ner, J. A. Melendez \textit{et al.}, Light nuclei with semilocal momentum-space regularized chiral interactions up to third order,
    { Phys. Rev. C {\bf 103}, 054001 (2021)}.
	
	
	\bibitem{Ragnar} S. R. Stroberg, H. Hergert, S. K. Bogner, and J. D. Holt, Nonempirical interactions for the nuclear shell model: An update,
    {Annu. Rev. Nucl. Part. Sci. 69, 307 (2019)}.
	
	
	\bibitem{CCEI} G. R. Jansen, M. D. Schuster,  A. Signoracci, G. Hagen, and P. Navratil, Open $sd$-shell nuclei from first principles, 
	{ Phys. Rev. C {\bf 94}, 011301(R) (2016)}.
	
	
	\bibitem{ab-initio2} K. Hebeler, J. D. Holt, J. Men\'endez, and A. Schwenk, Nuclear Forces and Their Impact on Neutron-Rich Nuclei and Neutron-Rich Matter,
	{Annu. Rev. Nucl. Part. Sci. \textbf{65}, 457 (2015).}
	
	\bibitem{ab-initio3} S. R. Stroberg, H. Hergert, S. K. Bogner, and J. D. Holt
	Nonempirical Interactions for the Nuclear Shell Model: An Update,
	{Annu. Rev. Nucl. Part. Sci. \textbf{69}, 307 (2019).}
	
	\bibitem{ab-initio1}W. Leidemann and G. Orlandini, Modern \textit{ab initio} approaches and applications in few-nucleon physics with A $\ge$ 4, 
	{Prog. Part. Nucl. Phys. {\bf 68}, 158 (2013).}
	
	\bibitem{CDB2K3} R. Machleidt, K. Holinde, and C. Elster, The Bonn meson-exchange model for the nucleon-nucleon interaction,
	{ Phys. Rep. {\bf 149}, 1 (1987).}
	
	
	
	
	\bibitem{QCD} R. Machleidt and D. R. Entem, Chiral effective field theory and nuclear forces,  
	{Phys. Rep. {\bf 503}, 1 (2011).}
	
	\bibitem{EFT1}E. Epelbaum, H.-W. Hammer, and U.-G. Mei\ss ner, Modern theory of nuclear forces, 
	{Rev. Mod. Phys. {\bf 81}, 1773 (2009).}
	
	\bibitem{EFT2}S. Weinberg, Phenomenological lagrangians, 
	{Physica A {\bf 96}, 327 (1979).}
	
	\bibitem{EFT3}S. Weinberg, Nuclear forces from chiral lagrangians, 
	{Phys. Lett. B {\bf 251}, 288 (1990).}
	
	\bibitem{EFT4}S. Weinberg, Effective chiral lagrangians for nucleon-pion interactions and nuclear forces, 
	{ Nucl. Phys. B {\bf 363}, 3 (1991).}
	
	
	
	
	\bibitem{NCSM_r1}P. Navr\'atil, S. Quaglioni, I. Stetcu, and B. R. Barrett, 
	Recent developments in no-core shell-model calculations,
	{ J. Phys. G: Nucl. Part. Phys. {\bf 36}, 083101 (2009).}
	

	
	
	
	\bibitem{SM1} B. A. Brown, The nuclear shell model towards the drip lines,
	{ Prog. Part. Nucl. Phys. {\bf 47}, 517 (2001).}
	
	\bibitem{SM2} S. Cohen and D. Kurath, Effective interactions for the $1p$ shell, 
	{Nucl. Phys. {\bf 73}, 1 (1965).}
	
	\bibitem{SM3} B. A. Brown and W. A. Richter, New “USD” Hamiltonians for the $sd$ shell, 
	{Phys. Rev. C {\bf 74}, 034315 (2006).}
	
	\bibitem{SM4} M. Honma, T. Otsuka, B. A. Brown, and T. Mizusaki, Shell-model description of neutron-rich $pf$-shell nuclei with a new effective interaction GXPF1,
	{Eur. Phys. J. A {\bf 25}, 499 (2005).}
	
	\bibitem{SM5} O. Sorlin and M.-G. Porquet,  Nuclear magic numbers: New features far from stability, 
	{Prog. Part. Nucl. Phys. {\bf 61}, 602 (2008).}
	
	\bibitem{SM6} T. Otsuka, A. Gade, O. Sorlin, T. Suzuki, and Y. Utsuno, Evolution of shell structure in exotic nuclei, 
	{Rev. Mod. Phys. {\bf 92}, 015002 (2020).}
	
	
	
	
	
	\bibitem{NCSM_p1}D. C. Zheng, J. P. Vary, and B. R. Barrett, Large-space shell-model calculations for light nuclei,
	{Phys. Rev. C {\bf 50}, 2841 (1994).}
	
	\bibitem{NCSM_p2}P. Navr\'atil and B. R. Barrett, No-core shell-model calculations with starting-energy-independent multivalued effective interactions, 
	{Phys. Rev. C {\bf 54}, 2986 (1996).}
	
	\bibitem{NCSM_p3}P. Navr\'atil and B. R. Barrett, Large-basis shell-model calculations for $p$-shell nuclei,
	{Phys. Rev. C {\bf 57}, 3119 (1998).}
	
 	\bibitem{NCSM_p4} P. Navr\'atil and B. R. Barrett,
 	Four-nucleon shell-model calculations in a Faddeev-like approach,
 	{ Phys. Rev. C {\bf 59}, 1906 (1999).}
%
	\bibitem{NCSM_p5} P. Navr\'atil, J. P. Vary, and B. R. Barrett, Properties of $^{12}$C in the \textit{ab initio} nuclear shell model, 
	{ Phys. Rev. Lett. {\bf 84}, 5728 (2000).}
	
	\bibitem{NCSM_p6} P. Navr\'atil, G. P. Kamuntavi\v{c}ius, and B. R. Barrett, 
	Few-nucleon systems in a translationally invariant harmonic oscillator basis, 
	{Phys. Rev. C {\bf 61}, 044001 (2000).}
	
	\bibitem{NCSM_p7} P. Navr\'atil, J. P. Vary, and B. R. Barrett, Large-basis \textit{ab initio} no-core shell model and its application to $^{12}$C,
	{Phys. Rev. C {\bf 62}, 054311 (2000).}
	
	\bibitem{NCSM_p8} P. Maris, J. P. Vary, and A. M. Shirokov, \textit{ Ab initio} no-core full configuration calculations of light nuclei, 
	{Phys. Rev. C {\bf 79}, 014308 (2009).}
	
	
	
	
	
	
	\bibitem{Ne0} Y. Z. Ma, F. R. Xu, N. Michel, S. Zhang, J. G. Li, B. S. Hu, L. Coraggio, N. Itaco, and
	A. Gargano, Continuum and three-nucleon force in Borromean system: The $^{17}$Ne
	case, 
	{Phys. Lett. B \textbf{808}, 135673 (2020).}
	
	\bibitem{Ne1} A. Arima, S. Cohen, R. D. Lawson, and M. H. MacFarlane, A shell-model study of the isotopes of O, F and Ne, 
	{Nucl. Phys. A \textbf{108}, 94 (1968).}
	
	\bibitem{Ne2} Y. Akiyama, A. Arima, and T. Sebe, The structure of the sd shell nuclei: (IV). $^{20}$Ne, $^{21}$Ne, $^{22}$Ne, $^{22}$Na and $^{24}$Mg, 
	{Nucl. Phys. A \textbf{138}, 273 (1969).}
	
	\bibitem{Ne2*} H. Sagawa, X. R. Zhou, X. Z. Zhang, and T. Suzuki, Deformations and electromagnetic moments in carbon and neon isotopes, 
	{Phys. Rev. C \textbf{70}, 054316 (2004).}
	
	\bibitem{Ne3} T. Tomoda and A. Arima, Coexistence of shell structure and cluster structure in $^{20}$Ne, 
	{Nucl. Phys. A \textbf{303}, 217 (1978).}
	
	\bibitem{Ne4} Y. Kanada-En'yo and H. Horiuchi, Coexistence of cluster and shell-model aspects in nuclear systems, 
	{Front. Phys. \textbf{13}, 132108 (2018).}
	
	\bibitem{Ne5} M. Frosini, T. Duguet, J.-P. Ebran, B. Bally, T. Mongelli, T. R. Rodríguez, R. Roth, and V. Somà, Multi-reference many-body perturbation theory for nuclei, 
	{Eur. Phys. J. A \textbf{58}, 63 (2022).}
	
	\bibitem{Ne7} T. Dytrych, K. D. Launey , J. P. Draayer , D. J. Rowe , J. L. Wood, G. Rosensteel, C. Bahri , D. Langr , and R. B. Baker, Physics of Nuclei: Key Role of an Emergent Symmetry, 
	{Phys. Rev. Lett. {\bf 124}, 042501 (2020).}
	
	\bibitem{Ne5*} O. M. Molchanov , K. D. Launey , A. Mercenne, G. H. Sargsyan , T. Dytrych, and J. P. Draayer, Machine learning approach to pattern recognition in nuclear dynamics from the $ab$ $initio$ symmetry-adapted no-core shell model, 
	{Phys. Rev. C \textbf{105}, 034306 (2022).}
	
	\bibitem{charge_radii2} T. Abe , P. Maris, T. Otsuka , N. Shimizu, Y. Utsuno, and J. P. Vary, Ground-state properties of light 4$n$ self-conjugate nuclei in $ab$ $initio$ no-core Monte Carlo shell model calculations with nonlocal $NN$ interactions, 
	{Phys. Rev. C {\bf 104}, 054315 (2021).}
	
	\bibitem{Ne5**} S. R. Stroberg, J. Henderson , G. Hackman, P. Ruotsalainen, G. Hagen,  and J. D. Holt, Systematics of $E2$ strength in the $sd$ shell with the valence-space in-medium similarity renormalization group, 
	{Phys. Rev. C \textbf{105}, 034333 (2022).}
	
	\bibitem{Ne6} P. Marević, J.-P. Ebran, E. Khan, T. Nikšić, and D. Vretenar, Quadrupole and octupole collectivity and cluster structures in neon isotopes, 
	{Phys. Rev. C {\bf 97}, 024334 (2018).}
	
	
	
	\bibitem{dripline1} M. Thoennessen, Current status and future potential of nuclide discoveries,
	{Rep. Prog. Phys. {\bf  76}, 056301 (2013).}
	
	\bibitem{dripline2}D. S. Ahn, N. Fukuda \textit{et al.}, Location of the neutron dripline at fluorine and neon, 
	{ Phys. Rev. Lett. {\bf 123}, 212501 (2019).}
	
	\bibitem{charge.rad1} \'A. Koszor\'us, X. F. Yang, W. G. Jiang $et$ $al.$,
	Charge radii of exotic potassium isotopes challenge nuclear theory and the magic character of N = 32,
	{Nat. Phys. \textbf{17}, 439 (2021)}.
	
	\bibitem{charge.rad2} I Angeli $et$ $al.$, $N$ and $Z$ dependence of nuclear charge radii, 
	{J. Phys. G: Nucl. Part. Phys. \textbf{36}, 085102 (2009)}.
	

	
	\bibitem{Ne10} W. Geithner, T. Neff  $et$ $al.$, Masses and Charge Radii of $^{17-22}$Ne and the Two-Proton-Halo Candidate $^{17}$Ne, 
	{Phys. Rev. Lett. \textbf{101}, 252502 (2008).}
	
	\bibitem{Ne9} K. Marinova, W. Geithner, M. Kowalska, K. Blaum, S. Kappertz, M. Keim, S. Kloos, G. Kotrotsios, P. Lievens, R. Neugart, H. Simon, and S. Wilbert, Charge radii of neon isotopes across the $sd$ neutron shell,
	{Phys. Rev. C {\bf 84}, 034313 (2011).}
	
	\bibitem{Ne8} S. J. Novario , G. Hagen , G. R. Jansen, and T. Papenbrock, Charge radii of exotic neon and magnesium isotopes, 
	{Phys. Rev. C {\bf 102}, 051303(R) (2020).}
	
	\bibitem{N=14} S. Bagchi, R. Kanungo, W. Horiuchi , G. Hagen, T. D. Morris, S. R. Stroberg, T. Suzuki, F. Ameil, J. Atkinson, Y. Ayyad \textit{et al.}, Neutron skin and signature of the $N=14$ shell gap found from measured proton radii of $^{17-22}$N, 
	{ Phys. Lett. B {\bf 790}, 251 (2019).} 
	

	
	
	
	
	\bibitem{INOY} P. Doleschall and I. Borb\'ely, Properties of the nonlocal $NN$ interactions required for the correct triton binding energy,
	{ Phys. Rev. C {\bf 62}, 054004 (2000).}
	
	\bibitem{CDBonn} R. Machleidt, High-precision, charge-dependent Bonn nucleon-nucleon potential, 
	{ Phys. Rev. C {\bf 63}, 024001 (2001).}
	
	\bibitem{N3LO} D. R. Entem and R. Machleidt, Accurate charge-dependent nucleon-nucleon potential at fourth order of chiral perturbation theory, 
	{ Phys. Rev. C {\bf 68}, 041001(R) (2003).}
	

	
	
	
	
	\bibitem{OLS1} S. \^{O}kubo,  Diagonalization of Hamiltonian and Tamm-Dancoff equation, 
	{ Prog. Theor. Phys. {\bf 12}, 603 (1954).}
	
	\bibitem{OLS2} K. Suzuki and S. Y. Lee, Convergent theory for effective interaction in nuclei, 
	{Prog. Theor. Phys. {\bf 64}, 2091 (1980).}
	
	\bibitem{OLS3} K. Suzuki, Construction of Hermitian effective interaction in nuclei: General relation between Hermitian and non-Hermitian forms, 
	{Prog. Theor. Phys. {\bf 68}, 246 (1982).}
	
	\bibitem{OLS4} K. Suzuki and R. Okamoto, Effective interaction theory and unitary-model-operator approach to nuclear saturation problem,
	{Prog. Theor. Phys. {\bf 92}, 1045 (1994).}
	
	
	
	
	
	\bibitem{SRG1} S. K. Bogner, R. J. Furnstahl, and R. J. Perry,
	Similarity renormalization group for nucleon-nucleon interactions,
	{Phys. Rev. C {\bf 75}, 061001 (2007).}
	
	\bibitem{SRG2} E. D. Jurgenson, P. Navr\'atil, and R. J. Furnstahl,
	Evolving nuclear many-body forces with the similarity renormalization group,
	{Phys. Rev. C {\bf 83}, 034301 (2011).}
	
	
	
	\bibitem{Lawson} D. H. Gloeckner and R. D. Lawson, Spurious center-of-mass motion, 
	{ Phys. Lett. B {\bf 53}, 313 (1974).}
	
	
	

	
	\bibitem{INOY2} P. Doleschall, I. Borb\'ely, Z. Papp, and W. Plessas, Nonlocality in the nucleon-nucleon interaction and three-nucleon bound states, 
	{ Phys. Rev. C {\bf 67}, 064005 (2003).}
	
	
	\bibitem{INOY3} P. Doleschall, Influence of the short range nonlocal nucleon-nucleon interaction on the elastic $n\ensuremath{-}d$ scattering: Below $30\phantom{\rule{0.3em}{0ex}}\mathrm{MeV}$,
	{Phys. Rev. C {\bf 69}, 054001 (2004).}
	

	
	
	\bibitem{av18}R. B. Wiringa, V. G. J. Stoks, and R. Schiavilla, Accurate nucleon-nucleon potential with charge-independence breaking, 
	{Phys. Rev. C {\bf 51},  38 (1995).}
	
	
	
	\bibitem{pAntoine1}E. Caurier and F. Nowacki, Present status of shell model techniques,
	{Acta Phys. Pol. B {\bf 30}, 705 (1999).}
	
	\bibitem{pAntoine2} E. Caurier, G. Mart\'{\i}nez-Pinedo, F. Nowacki, A. Poves, and A. P. Zuker, The shell model as a unified view of nuclear structure, 
	{ Rev. Mod. Phys. {\bf 77}, 427 (2005).}
	
	\bibitem{pAntoine3}C. Forss\'en, B. D. Carlsson, H. T. Johansson, D. S\"{a}\"{a}f, A. Bansal, G. Hagen, and T. Papenbrock, Large-scale exact diagonalizations reveal low-momentum scales of nuclei, 
	{ Phys. Rev. C {\bf 97}, 034328 (2018).}
	
	
	
	\bibitem{arch1} A. Saxena and P. C. Srivastava, \textit{Ab initio} no-core shell model study of neutron-rich nitrogen isotopes, 
	{Prog. Theor. Exp. Phys. {\bf 2019}, 073D02 (2019).}
	
	\bibitem{pr1} P. Choudhary, P. C. Srivastava, M. Gennari, P. Navr\'atil, \textit{Ab initio} no-core shell-model description of $^{10-14}$C isotopes, Phys. Rev. C  {\bf 117}, 014309 (2023).
	
	
	\bibitem{pr2} P. Choudhary and P. C. Srivastava, \textit{Ab initio}  no-core shell model study of neutron-rich $^{18,19,20}$C isotopes, Nucl. Phys. A {\bf 1029}, 122565 (2023).
	
	
	\bibitem{arch2} A. Saxena and P. C. Srivastava, \textit{Ab initio} no-core shell model study of $^{18-23}$O and $^{18-24}$F isotopes, 
	{J. Phys. G: Nucl. Part. Phys. {\bf 47}, 055113 (2020).}
	
	\bibitem{priyanka} P. Choudhary, P. C. Srivastava , and P. Navr\'atil, $Ab$ $initio$ no-core shell model study of $^{10-14}$B
	isotopes with realistic $NN$ interactions, 
	{Phys. Rev. C {\bf 102}, 044309 (2020).}
	
	

	\bibitem{Ragnar93}
	S. R. Stroberg, H. Hergert, J. D. Holt, S. K. Bogner, and A. Schwenk, Ground and excited states of doubly open-shell nuclei from ab initio valence-space Hamiltonians,
	{Phys. Rev. C {\bf 93}, 051301(R) (2016).}
	
	
	\bibitem{Jensen94}
	G. R. Jansen, M. D. Schuster, A. Signoracci, G. Hagen, and P. Navrátil,
	Open sd-shell nuclei from first principles, 
		{Phys. Rev. C {\bf 94}, 011301(R) (2016).}
	
	\bibitem{Jonathan102}
		J. Williams et al., High-spin structure of the sd shell nuclei $^{25}$Na and $^{22}$Ne,
	{Phys. Rev. C {\bf 102}, 064302 (2020).}
	
	
	
	\bibitem{NNDC}Data extracted using the NNDC World Wide Web site from the ENSDF, 
	{ https://www.nndc.bnl.gov/ensdf/.}	
	
	\bibitem{Qandmag}IAEA, 
	{https://www-nds.iaea.org/nuclearmoments/.}
	
	 	\bibitem{CDB2K2} R. Machleidt, F. Sammarruca, and Y. Song, Nonlocal nature of the nuclear force and its impact on nuclear structure, 
 	{ Phys. Rev. C {\bf 53},  R1483(R) (1996).}
	
	

	
	\bibitem{Roth-2016} A. Calci and R. Roth, Sensitivities and correlations of nuclear structure observables emerging from chiral interactions, {Phys. Rev. C {\bf 94}, 014322 (2016).}


	
	
	
	\bibitem{charge_radii} I. Angeli $et$ $al.$, Table of experimental nuclear g.s charge radii: An update, {At. Data Nucl. Data Tables, \textbf{99}, 69 (2013).}
	
	\bibitem{C.Forseen} A. Ekström, G. R. Jansen, K. A. Wendt, G. Hagen, T. Papenbrock, B. D. Carlsson,  C. Forssén, M. Hjorth-Jensen, P. Navrátil, and W. Nazarewicz, Accurate nuclear radii and binding energies from a chiral interaction, {Phys. Rev. C {\bf 91}, 051301(R) (2015).}
	
	\bibitem{charge_radii4} A. M. Shirokov, I. J. Shin, Y. Kim, M. Sosonkina, P. Maris, and J. P. Vary, N3LO $NN$ interaction adjusted to light nuclei in $ab$ $exitu$ approach, {Phys. Lett. B {\bf 761}, 87 (2016).}
	
	\bibitem{charge_radii3} M. A. Caprio, P. Maris, and J. P. Vary,  Halo nuclei $^6$He and $^8$He with the Coulomb-Sturmian basis, {Phys. Rev. C {\bf 90}, 034305 (2014).}
	
	\bibitem{Roth-2009} R. Roth, Importance truncation for large-scale configuration interaction approaches, {Phys. Rev. C {\bf 79}, 064324 (2009).}
	
	\bibitem{SANCSM_2206}G. H. Sargsyan, K. D. Launey, and M. T. Burkey, $et$ $al.$, Impact of clustering on the $^{8}$Li $\beta$ decay and recoil form factors, {Phys. Rev. Lett. {\bf 128}, 202503 (2022).}
	
	
	\bibitem{M.Caprio2019} S. L. Henderson, T. Ahn, M. A. Caprio $et$ $al.$, First measurement of the $B(E2; 3/2^- \to 1/2^-)$ transition strength in $^7$Be: Testing \textit{ab initio} predictions for $A$ = 7 nuclei, {Phys. Rev. C {\bf 99}, 064320 (2019).}
	
	\bibitem{M.Caprio2022} M. A. Caprio and P. J. Fasano, \textit{Ab initio} estimation of $E2$ strengths in $^{8}\mathrm{Li}$ and its neighbors by normalization to the measured quadrupole moment, {Phys. Rev. C {\bf 106}, 034320 (2022).}
	
	\bibitem{M.Caprio2022b} M. A. Caprio and P. J. Fasano, Robust \textit{ab initio} prediction of nuclear electric quadrupole
	observables by scaling to the charge radius, {Phys. Rev. C {\bf 105}, L061302 (2022).}
	

	
	
\end{thebibliography}
\end{document}